\documentclass[10pt,article]{JHEP3}
\usepackage{amsmath}
\usepackage{graphicx}
\usepackage{psfrag}
\usepackage{epsfig}

\title{$k-$Essence, superluminal propagation, causality and emergent geometry}

\author{Eugeny Babichev\\
INFN - Laboratori Nazionali del Gran Sasso, S.S. 17bis, 67010
Assergi (L'Aquila) - Italy, \\
Institute for Nuclear Research of RAS, 60th October Anniversary
Prospect 7a, 117312 Moscow, Russia
\\E-mail: \email{eugeny.babichev@lngs.infn.it}}

\author{Viatcheslav Mukhanov\\
Arnold-Sommerfeld-Center for Theoretical Physics, Department f\"ur Physik, 
Ludwig-Maximilians-Universit\"at M\"unchen,
Theresienstr. 37, D-80333, Munich, Germany, \\
E-mail: \email{mukhanov@theorie.physik.uni-muenchen.de}} 

\author{Alexander Vikman\\
Arnold-Sommerfeld-Center for Theoretical Physics, Department f\"ur Physik, 
Ludwig-Maximilians-Universit\"at M\"unchen,
Theresienstr. 37, D-80333, Munich, Germany
\\E-mail: \email{vikman@theorie.physik.uni-muenchen.de}} 

\preprint{LMU-ASC 54/07}
 
\abstract{The k-\emph{essence} theories admit in general the superluminal propagation
of the perturbations on classical backgrounds. We show that in spite
of the superluminal propagation the causal paradoxes do not arise
in these theories and in this respect they are not less safe than
General Relativity.}

\begin{document}

\section{Introduction}

Over the past years, spontaneous breaking of the Lorentz invariance
and questions related to this issue, such as superluminal propagation
of perturbations in nontrivial backgrounds, attracted a renewed interest
among physicists. One of the basic questions here is whether the theories
allowing the superluminal velocities possess internal inconsistencies
and, in particular, inevitably lead to the causality violation namely
to the appearance of the closed causal curves (CCCs). Concerning this
issue there exist two contradicting each other points of view. Some
authors (see, for instance, \cite{Hawking,Durrer,Ellis,Gibbons,Zwanziger,Susskind,Kleban,Nima})
argue that the subluminal propagation condition should a priori be
imposed to make the theory physically acceptable. For example, in~\cite{Hawking}
on the P. 60 the authors introduce the ``\emph{Postulate of Local
Causality}'' which excludes the superluminal velocities from the
very beginning. The requirement of subluminality is sometimes used
to impose rather strong restrictions on the form of the admissible
Lagrangians for the vector and higher spin fields \cite{Zwanziger}
and gravity modifications \cite{Kleban}. The effective field theories
(EFT) allowing the superluminal propagation were considered in \cite{Nima},
where it was argued that in such theories global causality and analyticity
of the S-matrix may be easily violated. The main conclusion of \cite{Nima}
is not favorable for the theories with superluminal propagation. In
particular the authors claim that the UV-completion of such theories
must be very nontrivial if it exists at all (for a different attitude
see \cite{Shore,Recent}).

An open minded opinion concerning the superluminal propagation is
expressed in \cite{Blochinzev}, where one argues that the proper
change of the chronological ordering of spacetime in non-linear field
theory with superluminal propagation allows us to avoid the causal
paradoxes.

Recently, in the literature were discussed several cases in which
faster-than-light propagation arises in a rather natural way. In particular
we would like to mention the noncommutative solitons \cite{Noncomutative},
Einstein aether waves \cite{Jacob}, ``superluminal'' photons
in the Drummond-Hathrell effect \cite{Faster than gravity,Ohkuma}
and in the Scharnhorst effect \cite{plates,UV} 
\footnote{In this paper under ``superluminal'' we
always mean ``faster than light in usual QED vacuum
in unbounded empty space''. To avoid confusion one
could say that photons propagate faster than gravitons in the Scharnhorst
effect. When this paper was in the final stage of preparation the
superluminal wave-front velocity in these effects was putted under
question \cite{Recent}.}. 
These last two phenomena are due to the vacuum polarization i.e.
higher-order QED corrections. It was argued that this superluminal
propagation leads to the causal paradoxes in the gedanken experiment
involving either two black holes \cite{Dolgov} or two pairs of Casimir
plates \cite{Liberati} moving with the high relative velocities.
To avoid the appearance of the closed causal curved in such experiments
the authors of~\cite{Liberati} invoked the \emph{Chronology Protection
Conjecture} \cite{Chronology} and showed that the photons in the
Scharnhorst effect causally propagate in effective metric different
from the Minkowski one.

Note that the superluminal propagation cannot be the sole reason for
the appearance of the closed causal curves. There are numerous examples
of spacetimes in General Relativity, where the \emph{``Postulate of Local Causality''} 
is satisfied and, nevertheless, the closed causal curves are present 
(see \cite{Godel,Gott,Ori,otherCCC,wormhole}).
Therefore an interesting question arises whether the superluminal
propagation leads to additional problems related with causality compared
to the situation in General Relativity.

In these paper we will consider the k-\emph{essence} fields \cite{Kessence,Super infl,GarMukh,BH}
and show that contrary to the claim of \cite{Durrer,Durrer2} the
causality is not violated in generic k-\emph{essence} models with
superluminal propagation (similar attitude was advocated in \cite{Bruneton,KimIrSen,Halo}).
In this sense, in spite of the presence of superluminal signals on
nontrivial backgrounds, the k-\emph{essence} theories are not less
safe and legitimate than General Relativity.

The paper is organized as follows. In Section \ref{SecII} we discuss the equation
of motion for k-\emph{essence} and derive generally covariant action
for perturbations for an arbitrary k-\emph{essence} background. 

General aspects of causality and propagation of perturbations on a
nontrivial background, determining the \textquotedblleft \emph{new
aether}\textquotedblright , are discussed in Section \ref{SecIII}. In particular,
we prove that no causal paradoxes arise in the cases studied in our
previous works \cite{Super infl,GarMukh,Kessence} and \cite{BH}.

Section \ref{SecIV} is devoted to the Cauchy problem for k-\emph{essence} equation
of motion. We investigate under which restrictions on the initial
conditions the Cauchy problem is well posed. 

In Section \ref{SecV} we study the Cauchy problem for small perturbations in
the \textquotedblleft \emph{new aether}\textquotedblright\ rest
frame and in the fast moving spacecraft.

Section \ref{SecVI} is devoted to the \emph{Chronology Protection Conjecture,}
which is used to avoid the CCCs in gedanken experiments considered
in \cite{Nima}. 

In Section \ref{SecVII} we discuss the universal role of the gravitational
metric. Namely, we show that for the physically justified k-\emph{essence}
theories the boundary of the smooth field configuration localized
in Minkowski vacuum, can propagate only with the speed not exceeding
the speed of light. In agreement with this result we derive that exact
solitary waves in purely kinetic k-\emph{essence} propagate in vacuum
with the speed of light. 

Our main conclusions are summarized in Section \ref{SecVIII}. 

All derivations of more technical nature the reader can find in Appendices.
In Appendix \ref{AppA} we derive characteristics of the equation of motion
and discuss local causality. Appendix \ref{AppB} is devoted to the derivation
of the generally covariant action for perturbations. In Appendix \ref{AppC}
we show how the action derived in Appendix \ref{AppB} is related to the action
for cosmological perturbations from \cite{GarMukh,Mukhanov BOOK}.
In Appendix \ref{AppD} we consider the connection between k-\emph{essence}
and hydrodynamics. The derivation of Green functions is given in Appendix \ref{AppE}.
\section{Equations of motion and emergent geometry}
\label{SecII}
Let us consider the k-\emph{essence} scalar field $\phi$ with the
action: \begin{equation}
S_{\phi}=\int d^{4}x\sqrt{-g}\mathcal{L}\left(X,\phi\right),\label{action}\end{equation}
where\[
X=\frac{1}{2}g^{\mu\nu}\nabla_{\mu}\phi\nabla_{\nu}\phi,\]
is the canonical kinetic term and by $\nabla_{\mu}$ we always denote
the covariant derivative associated with metric $g_{\mu\nu}$. We
would like to stress that this action is explicitly generally covariant
and Lorentz invariant. The variation of action (\ref{action}) with
respect to $g_{\mu\nu}$ gives us the following energy-momentum tensor
for the scalar field: \begin{equation}
T_{\mu\nu}\equiv\frac{2}{\sqrt{-g}}\frac{\delta S_{\phi}}{\delta g^{\mu\nu}}=\mathcal{L}_{,X}\nabla_{\mu}\phi\nabla_{\nu}\phi-g_{\mu\nu}\mathcal{L},\label{EMT}\end{equation}
where $\left(...\right)_{,X}$ is the partial derivative with respect
to $X$. The Null Energy Condition (NEC) $T_{\mu\nu}n^{\mu}n^{\nu}\geq0$
(where $n^{\mu}$ is a null vector: $g_{\mu\nu}n^{\mu}n^{\nu}=0$)
is satisfied provided $\mathcal{L}_{,X}\geq0$. Because violation
of this condition would imply the unbounded from below Hamiltonian
and hence signifies the inherent instability of the system \cite{Inst}
we consider only the theories with $\mathcal{L}_{,X}\geq0$.

The equation of motion for the scalar field is obtained by variation
of action (\ref{action}) with respect to $\phi$, \begin{equation}
-\frac{\delta S}{\delta\phi}=\tilde{G}^{\mu\nu}\nabla_{\mu}\nabla_{\nu}\phi+2X\mathcal{L}_{,X\phi}-\mathcal{L}_{,\phi}=0,\label{EOM}\end{equation}
where the \textquotedblleft effective\textquotedblright\ metric
is given by \begin{equation}
\tilde{G}^{\mu\nu}\left(\phi,\nabla\phi\right)\equiv\mathcal{L}_{,X}g^{\mu\nu}+\mathcal{L}_{,XX}\nabla^{\mu}\phi\nabla^{\nu}\phi.\label{Metric}\end{equation}
This second order differential equation is hyperbolic (that is, $\tilde{G}^{\mu\nu}$
has the Lorentzian signature) and hence describes the time evolution
of the system provided~\cite{Susskind,Rendall,Halo}\begin{equation}
1+2X\frac{\mathcal{L}_{,XX}}{\mathcal{L}_{,X}}>0.\label{HYPER}\end{equation}
When this condition holds everywhere the effective metric $\tilde{G}^{\mu\nu}$
determines the characteristics (cone of influence) for k-\emph{essence},
see e.g. \cite{Rendall,Halo,Novello,stringy causality}. For \emph{nontrivial
configurations} of k-\emph{essence} field $\partial_{\mu}\phi\neq0$
and the metric $\tilde{G}^{\mu\nu}$ is generally not conformally
equivalent to $g^{\mu\nu}$; hence in this case the characteristics
do not coincide with those ones for canonical scalar field the Lagrangian
of which depends linearly on the kinetic term $X$. In turn, the characteristics
determine the \emph{local causal structure} of the space time in every
point of the manifold. Hence, the \emph{local causal structure} for
the k-\emph{essence} field is generically different from those one
defined by metric $g_{\mu\nu}$ (see Appendix \ref{AppA} for details). For
the coupled system of equations for the gravitational field and k-\emph{essence}
the Cauchy problem is well posed only if the initial conditions are
posed on the hypersurface which is spacelike with respect to both
metrics: $g^{\mu\nu}$ and $\tilde{G}^{\mu\nu}$ (see P.~251 of Ref.~\cite{Wald}
and Refs.~\cite{Leray,Rendall,Cauchy} for details). We postpone
the detailed discussion of this issue until Section \ref{SecIV}
and now we turn to the behavior of small perturbations on a given
background.
With this purpose it is convenient to introduce the function
\begin{equation}
c_{s}^{2}\equiv\left(1+2X\frac{\mathcal{L}_{,XX}}{\mathcal{L}_{,X}}\right)^{-1},\label{cs}\end{equation}
which for the case $X>0$ plays the role of \textquotedblleft speed
of sound\textquotedblright\ for small perturbations \cite{GarMukh}
propagating in the preferred reference frame, where the background
is at rest. It is well known that in the case under consideration
there exists an equivalent hydrodynamic description of the system
(see Appendix \ref{AppD}) and the hyperbolicity condition (\ref{HYPER}) is
equivalent to the requirement of the hydrodynamic stability $c_{s}^{2}>0$.

The Leray's theorem (see P.~251 of Ref.~\cite{Wald} and Ref.~\cite{Leray}
) states that the perturbations $\pi$ on given background $\phi_{0}\left(x\right)$
propagate causally in metric $\tilde{G}^{\mu\nu}\left(\phi_{0},\nabla\phi_{0}\right)$.
In Appendix \ref{AppB} we show that neglecting the metric perturbations $\delta g_{\mu\nu},$
induced by $\pi,$ one can rewrite the equation of motion for the
scalar field perturbations in the following form \begin{equation}
\frac{1}{\sqrt{-G}}\partial_{\mu}\left(\sqrt{-G}G^{\mu\nu}\partial_{\nu}\pi\right)+M_{\text{eff}}^{2}\pi=0,\label{eompi}\end{equation}
here we denote\begin{eqnarray}
G^{\mu\nu}\equiv\frac{c_{s}}{\mathcal{L}_{,X}^{2}}\tilde{G}^{\mu\nu}, & \,\,\sqrt{-G}\equiv\sqrt{-\text{det}G_{\mu\nu}^{-1}} & \textrm{\,\, where\,\, }G_{\mu\lambda}^{-1}G^{\lambda\nu}=\delta_{\mu}^{\nu},\label{Gup}\end{eqnarray}
and \begin{equation}
M_{\text{eff}}^{2}\equiv\frac{c_{s}}{\mathcal{L}_{,X}^{2}}\left(2X\mathcal{L}_{,X\phi\phi}-\mathcal{L}_{,\phi\phi}+\frac{\partial\tilde{G}^{\mu\nu}}{\partial\phi}\nabla_{\mu}\nabla_{\nu}\phi_{0}\right).\label{Mass}\end{equation}
Note that the metric $G^{\mu\nu}$ is conformally equivalent to $\tilde{G}^{\mu\nu}$
and hence describes the same causal structure as it must be. The equation
for the perturbations has exactly the same form as equation for the
massive Klein-Gordon field in the curved spacetime. Therefore the
metric $G^{\mu\nu}$ describes the \textquotedblleft \emph{emergent}\textquotedblright\ or
\textquotedblleft \emph{analogue}\textquotedblright\ spacetime
where the perturbations live. In particular this means that the action
for perturbations\begin{equation}
S_{\pi}=\frac{1}{2}\int d^{4}x\sqrt{-G}\,\left[G^{\mu\nu}\partial_{\mu}\pi\partial_{\nu}\pi-M_{\text{eff}}^{2}\pi^{2}\right],\label{Spi}\end{equation}
 and the equation of motion (\ref{eompi}) are generally covariant
in the geometry $G^{\mu\nu}$. Introducing the covariant derivatives
$D_{\mu}$ associated with metric $G^{\mu\nu}$ $(D_{\mu}G^{\alpha\beta}=0)$,
equation (\ref{eompi}) becomes \begin{equation}
G^{\mu\nu}D_{\mu}D_{\nu}\pi+M_{\text{eff}}^{2}\pi=0.\label{EOM pi general}\end{equation}
Using the inverse to $G^{\mu\nu}$ matrix \begin{equation}
G_{\mu\nu}^{-1}=\frac{\mathcal{L}_{,X}}{c_{s}}\left[g_{\mu\nu}-c_{s}^{2}\left(\frac{\mathcal{L}_{,XX}}{\mathcal{L}_{,X}}\right)\nabla_{\mu}\phi_{0}\nabla_{\nu}\phi_{0}\right],\label{Gdown}\end{equation}
one can define the \textquotedblleft emergent\textquotedblright\ interval
\begin{equation}
dS^{2}\equiv G_{\mu\nu}^{-1}dx^{\mu}dx^{\nu},\label{interval}\end{equation}
which determines the influence cone for small perturbations of k-\emph{essence}
on a given background%
\footnote{Note that in order to avoid confusion we will be raising and lowering
the indices of tensors by gravitational metric $g^{\mu\nu}$ ($g_{\mu\nu}$)
throughout the paper.%
}. This influence cone is larger than those one determined by the metric
$g_{\mu\nu},$ provided $\mathcal{L}_{,XX}/\mathcal{L}_{,X}<0$~\cite{Susskind,Rendall,Halo,stringy causality},
and the superluminal propagation of small perturbations becomes possible
(see Appendix \ref{AppA}). At first glance it looks like the theory under consideration
has emergent bimetric structure. However, this theory is inherently
different from the bimetric theories of gravity \cite{BI} because
the emergent metric refers only to the perturbations of k-\emph{essence}
and is due to the non-linearity of the theory, while in the bimetric
gravity theories both metrics have fundamental origin and are on the
same footing. 

The derived above form of the action and of the equation of motion
for perturbations is very useful. In particular, it simplifies the
stability analysis of the background with respect to the perturbations
of arbitrary wavelengths, while the hyperbolicity condition (\ref{HYPER})
guarantees this stability only with respect to the short-wavelength
perturbations. 

It is important to mention that besides of the usual hyperbolicity
condition (\ref{HYPER}) one has to require that $\mathcal{L}_{,X}$
is nowhere vanishes or becomes infinite. The points where $\mathcal{L}_{,X}$
vanishes or diverges, generally correspond to the singularities of
the emergent geometry. It follows from equations~(\ref{Gup}) and
(\ref{Gdown}) that these singularities are of the true nature and
cannot be avoided by the change of the coordinate system. Therefore
one can argue that before the singularities are formed the curvature
of the emergent spacetime becomes large enough for efficient quantum
production of the k-\emph{essence} perturbations which will destroy
the classical background and therefore $\mathcal{L}_{,X}$ cannot
dynamically change its sign. Hence, if one assumes that at some moment
of time the k-\emph{essence} satisfies the null energy condition,
that is, $\mathcal{L}_{,X}>0$ everywhere in the space (or $\varepsilon+p>0$
in hydrodynamical language; see Appendix \ref{AppD}) then this condition can
be violated only if one finds the way to pass through the singularity
in the emergent geometry with taking into account the quantum production
of the perturbations. This doubts the possibility of the smooth crossing
of the equation of state $w=-1$ and puts under question recently
suggested models of the bouncing universe (\cite{Bounce}). The statements
above generalize the results obtained in \cite{Ja} and re-derived
later in different ways in \cite{Crossing} in cosmological context.

In deriving (\ref{Spi}) and (\ref{EOM pi general}) we have assumed
that the k-\emph{essence} is sub-dominant component in producing the
gravitational field and consequently have neglected the metric perturbations
induced by the scalar field. In particular the formalism developed
is applicable for accretion of a \emph{test} scalar field onto black
hole \cite{BH}. For k-\emph{essence} dark energy \cite{Kessence}
action (\ref{Spi}) can be used only when k-\emph{essence} is a small
fraction of the total energy density of the universe, in particular,
this action is applicable during the stage when the speed of sound
of a successful k-\emph{essence} has to be larger than the speed of
light \cite{Durrer,KimIrSen}. During k-\emph{inflation} \cite{kinfl,kself,Super infl}
or DBI inflation \cite{DBI} the geometry $g_{\mu\nu}$ is determined
by the scalar field itself and therefore the induced scalar metric
perturbations are of the same order of magnitude as the perturbations
of the scalar field. For this case the action for cosmological perturbations
was derived in \cite{GarMukh}, see also \cite{Mukhanov BOOK}. We
have shown in Appendix \ref{AppC} that the correct action for perturbations
in k-\emph{inflation} has, however, the same structure of the kinetic
terms as (\ref{Spi}) or, in other words, the perturbations live in
the same emergent spacetime with geometry $G^{\mu\nu}$. One can expect
therefore that this emergent geometry $G^{\mu\nu}$ has a much broader
range of applicability and determines the causal structure for perturbations
also in the case of other backgrounds, where one cannot neglect the
induced metric perturbations.

If the hyperbolicity condition (\ref{HYPER}) is satisfied, then at
any given point of spacetime the metric $G_{\mu\nu}^{-1}$ can always
be brought to the canonical Minkowski form $\text{diag}\left(1,-1,-1,-1\right)$
by the appropriate coordinate transformation. However, the quadratic
forms $g_{\mu\nu}$ and $G_{\mu\nu}^{-1}$ are not positively defined
and therefore for a general background there exist no coordinate system
where they are both simultaneously diagonal. In some cases both metrics
can be nevertheless simultaneously diagonalized at a given point,
so that, e.g. gravitational metric $g_{\mu\nu}$ is equal Minkowski
metric and the induced metric $G_{\mu\nu}^{-1}$ is proportional to
$\text{diag}\left(c_{s}^{2},-1,-1,-1\right),$ where $c_{s}$ is the
speed of sound (\ref{cs}). For instance, in isotropic homogeneous
universe both metrics are always diagonal in the Friedmann coordinate
frame.

We conclude this section with the following interesting observation.
The effective metric (\ref{Gdown}) can be expressed through the energy
momentum tensor (\ref{EMT}) as
\begin{equation}
G_{\mu\nu}^{-1}=\alpha g_{\mu\nu}+\beta T_{\mu\nu}\end{equation}
where \[
\alpha=\frac{\mathcal{L}_{,X}}{c_{s}}-\mathcal{L}c_{s}\frac{\mathcal{L}_{,XX}}{\mathcal{L}_{,X}}\text{\,\,\, and\,\,\,}\beta=-c_{s}\frac{\mathcal{L}_{,XX}}{\mathcal{L}_{,X}}.\]
As we have pointed out the cosmological perturbations propagate in
$G_{\mu\nu}^{-1}$ even if the background field determines the dynamics
of the universe. In this case the energy momentum tensor for the scalar
filed satisfies the Einstein equations and eventually we can rewrite
the effective metric in the following form
\begin{equation}
G_{\mu\nu}^{-1}=\left(\alpha-\frac{\beta}{2}R\right)g_{\mu\nu}+\beta R_{\mu\nu}.\label{Redefinition}
\end{equation}
This looks very similar to the \textquotedblleft metric redefinition\textquotedblright\ $g_{\mu\nu}\leftrightarrow G_{\mu\nu}^{-1}$
in string theory where the quadratic in curvature terms in the effective
action are fixed only up to \textquotedblleft metric redefinition\textquotedblright\ (\ref{Redefinition})
see e.g. \cite{Metric Redefinition}. The \textquotedblleft metric
redefinition\textquotedblright\ does not change the light cone and
hence the \emph{local causality} only in the Ricci flat $R_{\mu\nu}=0$
spacetimes. However, neither in the matter dominated universe nor
during inflation the \emph{local causals structures} determined by
$g_{\mu\nu}$ and $G_{\mu\nu}^{-1}$ are equivalent. 
\section{Causality on nontrivial backgrounds}
\label{SecIII}
\FIGURE{\label{Bad Rocket}
\psfrag{T}[l]{$t$} 
\psfrag{X}[b]{$x$}
\psfrag{tt}[b]{$t'$}
\psfrag{XX}[b]{$x'$}
\epsfig{file=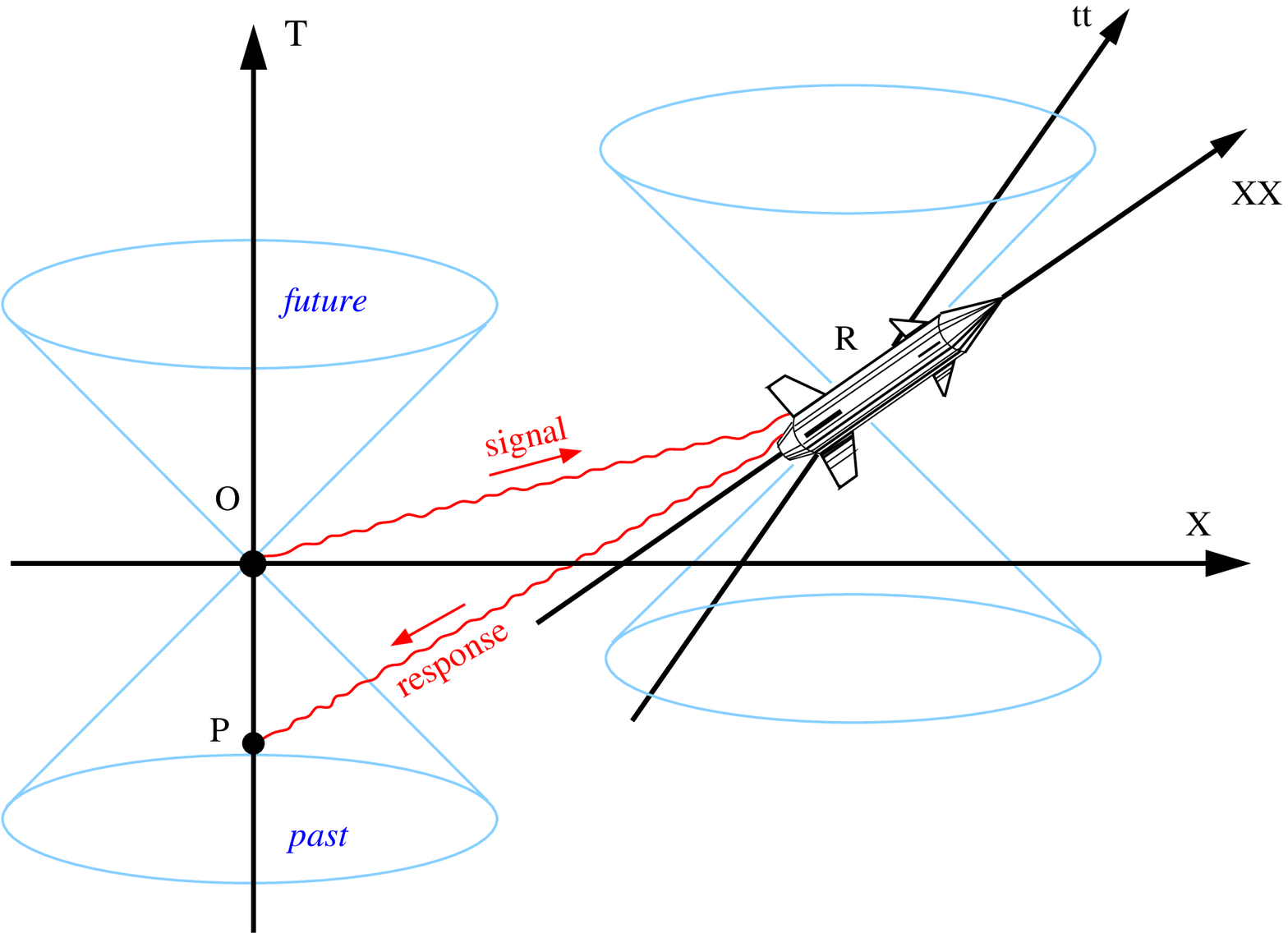,width=14cm,height=10cm}
\caption{ 
This figure represents the causal paradox constructed
using \emph{tachyons}. Someone living along the worldline $x=0$ sends
a tachyon signal to the astronaut in a fast moving spacecraft, $OR$.
In the spacecraft frame $(x',t')$, the astronaut sends a tachyon
signal back, $RP$. The signal $RP$ propagates in the direction of
growing $t'$ as it is seen by the astronaut, however it travels \emph{``back
in time''} in the rest frame. Thus it is possible to send a message
back in the own past.}
}
\FIGURE{\label{Good Rocket}
  \psfrag{T}[l]{$\large{t}$} 
  \psfrag{X}[b]{$x$}
  \psfrag{tt}[b]{$t'$}
  \psfrag{XX}[b]{$x'$}
\epsfig{file=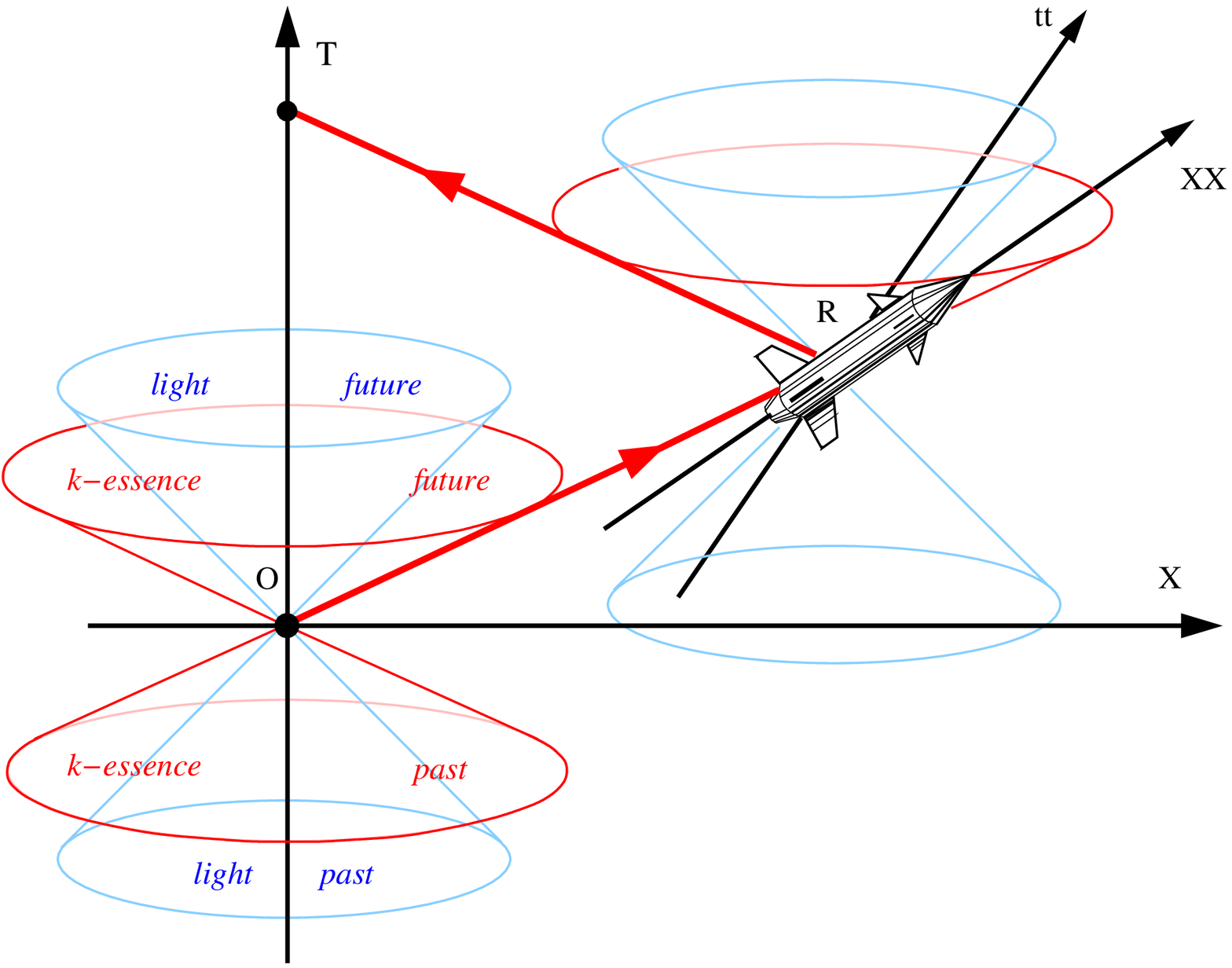,width=14cm,height=10cm}
\caption{
The causality paradox is avoided when superluminal signals propagate
in the background which breaks the Lorentz symmetry (compare with
Fig.~\ref{Bad Rocket}). The observers cannot send a message to themselves
in the past.}
}
In this section we discuss the causality issue for superluminal propagation
of perturbations on some nontrivial backgrounds, in particular, in
Minkowski spacetime with the scalar field, in Friedmann universe and
for black hole surrounded by the accreting scalar field.

First, we would like to recall a well-known paradox sometimes called
\textquotedblleft tachyonic anti-telephone\textquotedblright\ \cite{Tolman}
arising in the presence of the superluminal hypothetical particles
\emph{tachyons} possessing unbounded velocity $c_{tachyon}>1$. In
this case we could send a message to our own past. Indeed, let us
consider some observer, who is at rest at $x=0$ with respect to the
reference frame $\left(x,t\right)$ and sends along $OR$ a tachyon
signal to an astronaut in the spacecraft $R$ (see Fig.~\ref{Bad Rocket}).
In turn, after receiving this signal, the astronaut communicates back
sending the tachyon signal, $RP$. As this signal propagates the astronaut
proper time $t^{\prime}$ grows. However, if the speed of the spacecraft
is larger than $1/c_{tachyon}$, then the signal $RP$ propagates
\emph{backward in time} in the original rest frame of the observer.
Thus, the observers can in principle send information from \textquotedblleft their
future\textquotedblright\ to \textquotedblleft their past\textquotedblright .
It is clear that such situation is unacceptable from the physical
point of view.

Now let us turn to the case of the Minkowski space-time filled with
the scalar field, which allows the \textquotedblleft superluminal\textquotedblright\
propagation of perturbations in its background. For simplicity we
consider a homogeneous time dependant field $\phi_{0}\left(t\right)$.
Its \textquotedblleft velocity\textquotedblright\  $\partial_{\mu}\phi$
is directed along the timelike vector, $u^{\mu}=(1,0,0,0)$. Why does
the paradox above not arise here? This is because the superluminal
propagation of the signals is possible only in the presence of nontrivial
background of scalar field which serves as the \emph{aether} for sonic
perturbations. The \emph{aether} selects the preferred reference frame
and clearly the equation of motion for acoustic perturbations is not
invariant under the Lorentz transformations unless $c_{s}=1$. In
the moving frame of the astronaut the equation for perturbations has
more complicated form than in the rest frame and the analysis of its
solutions is more involved. However, keeping in mind that k-\emph{essence}
signals propagate along the characteristics which are coordinate independent
hypersurfaces in the spacetime we can study the propagation of sonic
perturbations, caused by the astronaut, in the rest frame of the aether
and easily find that the signal propagates always \emph{forward in
time} in this frame (see Fig.~\ref{Good Rocket}). Hence no closed
causal curves can arise here.

We would like to make a remark concerning the notion of \textquotedblleft
future-\textquotedblright\ and \textquotedblleft past\textquotedblright\
directed signals. It was argued in \ \cite{Durrer2} that in order
to have no CCCs for the k-\emph{essence} during the \textquotedblleft
superluminal\textquotedblright\ stage, \textquotedblleft ...the
observers travelling at high speeds with respect to the cosmological
frame must send signals backwards in their time for some specific
direction\textquotedblright . One should remember, however, that
the notion of past and future is determined by the past and future
cones in the spacetime and has nothing to do with a particular choice
of coordinates. Thus, the signals, which are future-directed in the
rest-frame remain the future-directed also in a fast-moving spacecraft,
in spite of the fact that this would correspond to the decreasing
time coordinate $t^{\prime}$. As we show in Section~\ref{SecV},
the confusion arises because of a poor choice of coordinates, when
decreasing $t^{\prime}$ correspond to future-directed signals and
vice versa. The example shown in Fig.~\ref{Figure Fluid} illustrates
this point: one can see that even without involving superluminal signals,
an increasing coordinate time does not always imply the future direction.

Another potentially confusing issue is related to the question which
particular velocity must be associated with the speed of signal propagation,
namely, phase, group or front velocity. For example, in \cite{Durrer2}
an acausal paradox is designed using different superluminal group
velocities for different wavenumbers. One should remember, however,
that neither group nor phase velocities have any direct relation with
the causal structure of the spacetime. Indeed the characteristic surfaces
of the partial differential equations describe the propagation of
the wavefront. This front velocity coincides with the phase velocity
only in the limit of the short wavelength perturbations. Generally
the wavefront corresponds to the discontinuity of the second derivatives
and therefore it moves \textquotedblleft off-shell\textquotedblright\ (a
more detailed discussion can be found in e.g.~\cite{Shore}). The
group velocity can be less or even larger than the wavefront velocity.
One can recall the simple examples of the canonical free scalar field
theories: for normal scalar fields the mass squared, $m^{2}>0,$ is
positive and the phase velocity is larger than $c$ while the group
velocity is smaller than $c$; on the other hand for tachyons ($m^{2}<0$)
the situation is opposite. Thus, if the group velocity were the speed
of the signal transfer, one could easily build the time-machine similar
to those described in \cite{Durrer2} using canonical scalar field
with negative mass squared, $m^{2}<0$. This, however, is impossible
because the causal structure in both cases ($m^{2}>0$ and $m^{2}<0$)
is governed by the same \emph{light} cones. Finally we would like
to mention that the faster-than-light group velocity has been already
measured in the experiment \cite{SuperGroup}. 

To prove the absence of the closed causal curves (CCC) in those known
situations where the superluminal propagation is possible, we use
the theorem from Ref.~\cite{Wald} (see p. 198): \emph{A spacetime
$\left(\mathcal{M},g_{\mu\nu}\right)$ is stably causal if and only
if there exists a differentiable function $f$ on $\mathcal{M}$ such
that $\nabla^{\mu}f$ is a future directed timelike vector field}.
Here $\mathcal{M}$ is a manifold and $g_{\mu\nu}$ is metric with
Lorentzian signature. Note, that the notion of \emph{stable causality}
implies that the spacetime $\left(\mathcal{M},\, g_{\mu\nu}\right)$
possesses no CCCs and thus no causal paradoxes can arise in this case.
The theorem above has a kinematic origin and does not rely on the
dynamical equations. In the case of the effective acoustic geometry
the acoustic metric $G_{\mu\nu}^{-1}$ plays the role of $g_{\mu\nu}$
and the function $f$ serves as the \textquotedblleft global time
function\textquotedblright\ of the emergent spacetime $\left(\mathcal{M},\, G_{\mu\nu}^{-1}\right)$.
For example, in the Minkowski spacetime filled with the scalar field
\textquotedblleft
ether\textquotedblright\ one can take the Minkowski time $t$ of
the rest frame, where this field is homogeneous, as the global time
function. Then we have 
\begin{equation}
G^{\mu\nu}\partial_{\mu}t\partial_{\nu}t=\frac{c_{s}}{\mathcal{L}_{,X}}g^{00}\left(1+2X\frac{\mathcal{L}_{,XX}}{\mathcal{L}_{,X}}\right)=\frac{g^{00}}{\mathcal{L}_{,X}c_{s}}.\label{Minkowski theorem}
\end{equation}
Even for those cases when the speed of perturbations can exceed the
speed of light, $c_{s}>1,$ this expression is positive, provided
that $\mathcal{L}_{,X}>0$, and the hyperbolicity condition (\ref{HYPER})
is satisfied. Thus $\partial_{\mu}t$ is timelike with respect to
the effective metric $G_{\mu\nu}^{-1}$; hence the conditions of the
theorem above are met and no CCCs can exist.

Now we consider the Minkowski spacetime with an arbitrary inhomogeneous
background $\phi_{0}\left(x\right)$ and verify under which conditions
one can find a global time $t$ for both geometries $g_{\mu\nu}$
and $G_{\mu\nu}^{-1}$ and thus guarantee the absence of CCCs. Let
us take the Minkowski $t$, $\eta^{\mu\nu}\partial_{\mu}t\partial_{\nu}t=1,$
and check whether this time can also be used as a global time for
$G_{\mu\nu}^{-1}$. We have \begin{equation}
G^{\mu\nu}\partial_{\mu}t\partial_{\nu}t=\frac{c_{s}}{\mathcal{L}_{,X}}\left[1+\left(\frac{\mathcal{L}_{,XX}}{\mathcal{L}_{,X}}\right)\left(\partial_{\mu}t\nabla^{\mu}\phi_{0}\right)^{2}\right]=\frac{c_{s}}{\mathcal{L}_{,X}}\left[1+\left(\frac{\mathcal{L}_{,XX}}{\mathcal{L}_{,X}}\right)\dot{\phi}_{0}^{2}\right],\end{equation}
and assuming that $c_{s}>0$, $\mathcal{L}_{,X}>0$ we arrive to the
conclusion that $t$ is a global time for emergent spacetime provided
\begin{equation}
1+\left(\frac{\mathcal{L}_{,XX}}{\mathcal{L}_{,X}}\right)\left(\dot{\phi}_{0}\left(x^{\mu}\right)\right)^{2}>0,\label{No CCCs Minkowski}\end{equation}
holds everywhere on the manifold $\mathcal{M}$. This inequality is
obviously always satisfied in the subluminal case. It can be rewritten
in the following form \begin{equation}
1+c_{s}^{2}\left(\frac{\mathcal{L}_{,XX}}{\mathcal{L}_{,X}}\right)\left(\vec{\nabla}\phi_{0}\left(x^{\mu}\right)\right)^{2}>0,\end{equation}
from where it is obvious that, if the spatial derivatives are sufficiently
small then this condition can also be satisfied even if $c_{s}>1$.
Note that the breaking of the above condition for some background
field configuration $\phi_{0}\left(x\right)$ does not automatically
mean the appearance of the CCCs. This just tells us that the time
coordinate $t$ cannot be used as the global time coordinate. However
it does not exclude the possibility that there exists another function
serving as the global time. Only, if one can prove that such global
time for both metrics does not exist at all, then there arise causal
paradoxes.

In the case of the Friedmann universe with \textquotedblleft
superluminal\textquotedblright\ scalar field, one can choose the
cosmological time $t$ as the global time function and then we again
arrive to~(\ref{Minkowski theorem}), thus concluding that there
exist no CCCs. In particular, the k-\emph{essence} models, where the
superluminal propagation is the generic property of the fluctuations
during some stage of expansion of the universe \cite{Durrer,KimIrSen},
\emph{do not lead to causal} \emph{paradoxes} contrary to the claim
by \cite{Durrer,Durrer2}.

The absence of the closed causal curves in the Friedmann universe
with k-\emph{essence} can also be seen directly by calculating of
the \textquotedblleft effective\textquotedblright\ line element
(\ref{interval}). Taking into account that the Friedmann metric is
given by \begin{equation}
ds^{2}=g_{\mu\nu}dx^{\mu}dx^{\nu}=dt^{2}-a^{2}(t)d\mathbf{x}^{2},\label{Friedmann line element}\end{equation}
we find that the line element (\ref{interval}), corresponding to
the effective acoustic metric, is \begin{equation}
dS^{2}=G_{\mu\nu}^{-1}dx^{\mu}dx^{\nu}=\frac{\mathcal{L}_{,X}}{c_{s}}\left(c_{s}^{2}dt^{2}-a^{2}(t)d\mathbf{x}^{2}\right).\label{effective line element}\end{equation}
The theory under consideration is generally covariant. After making
redefinitions, $\sqrt{\mathcal{L}_{,X}c_{s}}dt\rightarrow dt,$ and,
$a^{2}(t)\mathcal{L}_{,X}/c_{s}\rightarrow a^{2}(t),$ the line element
(\ref{effective line element}) reduces to the interval for the Friedmann
universe (\ref{Friedmann line element}), where obviously no causality
violation can occur. Thus we conclude that both the k-\emph{essence}
\cite{Kessence} and the \textquotedblleft superluminal\textquotedblright\ inflation
with large gravity waves \cite{Super infl} are completely safe and
legitimate on the side of causality.

When $X=\frac{1}{2}g^{\mu\nu}\partial_{\mu}\phi_{0}\partial_{\nu}\phi_{0}$
is positive everywhere in the spacetime the background field itself
can be used as the global time function. Indeed for general gravitational
background $g_{\mu\nu}$ and $c_{s}>0$, $\mathcal{L}_{,X}>0$ we
have \[
g^{\mu\nu}\partial_{\mu}\phi_{0}\partial_{\nu}\phi_{0}>0\text{ and }G^{\mu\nu}\partial_{\mu}\phi_{0}\partial_{\nu}\phi_{0}=\frac{2X}{\mathcal{L}_{,X}c_{s}}>0,\]
and due to the fact that $X>0$ the sign in front $\nabla^{\mu}\phi_{0}$
can be chosen so that the vector $\nabla^{\mu}\phi_{0}$ is always
future directed on $\mathcal{M}$. Therefore $\phi_{0}\left(x\right)$
or ($-\phi_{0}\left(x\right)$ if necessary) can serve as a global
time in both spacetimes $\left(\mathcal{M},\, g_{\mu\nu}\right)$
and $\left(\mathcal{M},\, G_{\mu\nu}^{-1}\right)$, and no causal
paradoxes arise.

In particular this is applicable for the accretion of the \textquotedblleft superluminal\textquotedblright\ scalar field onto the Schwarzschild black hole \cite{BH}. In this case sound
horizon is located inside the Schwarzschild radius and therefore the
Schwarzschild time coordinate cannot be used as a global time function.
However, $X>0$ outside the acoustic horizon (see \cite{BH}) and
in accordance with the theorem and discussion above we can take $\phi$
as the global time coordinate and hence the acoustic spacetime is
stably causal.

In all examples above we have considered the \textquotedblleft
superluminal\textquotedblright\ acoustic metric. Thus, if there
exist no CCCs in $\left(\mathcal{M},\, G_{\mu\nu}^{-1}\right)$ then
there are no CCCs with respect to metric $g_{\mu\nu}$ because acoustic
cone is larger than the light cone. It may happen that in some cases
it is not enough to prove that there no CCCs separately in $\left(\mathcal{M},\, G_{\mu\nu}^{-1}\right)$
and $\left(\mathcal{M},\, g_{\mu\nu}\right)$ and one has to use the
maximal cone or introduce an artificial cone \cite{Bruneton} encompassing
all cones arising in the problem. It is interesting to note that,
if the k\emph{-essence} realizes both \textquotedblleft
superluminal\textquotedblright\ and subluminal speed of sound in
the different regions of the manifold, then there exist hypersurface
where the k\emph{-essence} metric is conformally equivalent to the
$g_{\mu\nu}$ and one can smoothly glue the maximal cones together
everywhere on $\mathcal{M}$. After that one can consider a new \textquotedblleft artificial
metric\textquotedblright\ $G_{\mu\nu}^{\Sigma}$ as determining
the complete causal structure of the manifold.

We would like to point out that although the theorem on stable causality
allowed us to prove that there is no causal paradoxes in those cases
we considered above, it is no guaranteed that CCCs cannot arise for
some other backgrounds. Indeed, in \cite{Nima} the authors have found
some configurations of fields possessing CCCs: one for the scalar
field with non-canonical kinetic term and another for the \textquotedblleft
wrong\textquotedblright -signed Euler-Heisenberg system. In both
cases the small perturbations propagate superluminally on rather non-trivial
backgrounds. We will pursue this issue further in Section \ref{SecVI}.
\section{Which initial data are allowed for the well posed Cauchy problem?
\label{SecIV}}
Using the theorem on stable causality we have proven that the \textquotedblleft superluminal\textquotedblright\ k-\emph{essence} does not lead to any causal paradoxes for cosmological solutions and
for accretion onto black hole. However, the consideration above is
of a kinematic nature and it does not deal with the question how to
pose the Cauchy problem for the background field $\phi_{0}$ and it's
perturbations $\pi$.

It was pointed out in \cite{Nima} that in the reference frame
of the spacecraft moving with respect to nontrivial background, where
$c_{s}>1,$ with the speed $v=1/c_{s}$ the Cauchy problem for small
perturbations $\pi$ is ill posed. This happens because the hypersurface
of the constant proper time $t^{\prime}$ of the astronaut is a null-like
with respect to the acoustic metric $G_{\mu\nu}^{-1}$. Hence $t^{\prime}=const$
is tangential to the characteristic surface (or sonic cone see Appendix \ref{AppA}) and cannot be used to formulate the Cauchy problem for perturbations
which \textquotedblleft live\textquotedblright\ in this acoustic
metric. Intuitively this happens because the perturbations propagate
instantaneously with respect to the hypersurface $t^{\prime}=const$.
Moreover, for $v>1/c_{s}$, the sonic cone deeps below the surface
$t^{\prime}=const$ (see Figs.~\ref{Cauchy 3d} and \ref{Good Rocket})
and in the spacetimes of dimension $D>2$ the Cauchy problem is ill
posed as well because there always exist two directions along which
the perturbations propagate \textquotedblleft instantaneously\textquotedblright\ in
time $t^{\prime}$ (red vectors in Fig.~\ref{Cauchy 3d}). This tell
us that not every imaginable configuration of the background can be
realized as the result of evolution of the system with the well formulated
Cauchy problem and hence not every set of initial conditions for the
scalar field is allowed.

In this Section we will find under which restrictions on the initial
configuration of the scalar field the Cauchy problem for equation
(\ref{EOM}) is well-posed. For this purpose it is more convenient
not to split the scalar field into background and perturbations and
consider instead the total value of the field $\phi=\phi_{0}+\pi$.
The k-\emph{essence} field interacts with gravity and therefore for
consistency one has to consider the coupled system of equations for
the gravitational metric $g_{\mu\nu}$ and the k-\emph{essence} field
$\phi$. In this case the Cauchy problem is well posed only if the
initial data are set up on a hypersurface $\Sigma$ which is simultaneously
spacelike in both metrics: $g_{\mu\nu}$ and $G_{\mu\nu}^{-1}$ (for
details see P.~251 of Ref.~\cite{Wald} and Refs.~\cite{Leray},
\cite{Rendall}, \cite{Cauchy}). We will work in the synchronous
coordinate system, where the metric takes the form \begin{equation}
ds^{2}=dt^{2}-\gamma_{ik}dx^{i}dx^{k},\end{equation}
and select the spacelike in $g_{\mu\nu}$ hypersurface $\Sigma$ to
be a constant time hypersurface $t=t_{0}$. The 1-form $\partial_{\mu}t$
vanishes on any vector $R^{\mu}$ tangential to $\Sigma$: $R^{\mu}\partial_{\mu}t=0$
(see Fig.~\ref{Cauchy 3d}). This 1-form is timelike with respect
to the gravitational metric $g_{\mu\nu}$, that is $g^{\mu\nu}\partial_{\mu}t\partial_{\nu}t>0$.
In case when Lagrangian for k-\emph{essence} depends at maximum on
the first derivatives of scalar field the initial conditions which
completely specify the unambiguous solution of the equations of motion
are the initial field configuration $\phi(\mathbf{x})$ and and it's
first time derivative $\dot{\phi}(\mathbf{x})\equiv\left(g^{\mu\nu}\partial_{\mu}t\partial_{\nu}\phi\right)_{\Sigma}$.
Given these initial conditions one can calculate the metric $G_{\mu\nu}^{-1}$
and consequently the influence cone at every point on $\Sigma$. First
we have to require that for a given set of initial data the hyperbolicity
condition (\ref{HYPER}) is not violated. This imposes the following
restriction on the allowed initial values $\,\,\phi(\mathbf{x})$
and $\dot{\phi}(\mathbf{x})$: \begin{equation}
c_{s}^{-2}=1+\left[\left(\dot{\phi}(\mathbf{x})\right)^{2}-\left(\vec{\nabla}\phi(\mathbf{x})\right)^{2}\right]\frac{\mathcal{L}_{,XX}}{\mathcal{L}_{,X}}>0,\label{hyper initial}\end{equation}
 where we have denoted $\left(\vec{\nabla}\phi(\mathbf{x})\right)^{2}=\gamma^{ik}\partial_{i}\phi\partial_{k}\phi$.
In addition we have to require that the hypersurface $\Sigma$ is
spacelike also with respect to emergent metric $G^{\mu\nu}$, that
is, for every vector $R^{\mu},$ tangential to $\Sigma,$ we have
$G_{\mu\nu}^{-1}R^{\mu}R^{\nu}<0,$ or \begin{equation}
1+c_{s}^{2}\left(\vec{\nabla}\phi(\mathbf{x})\right)^{2}\frac{\mathcal{L}_{,XX}}{\mathcal{L}_{,X}}>0.\label{NO dips}\end{equation}
If at some point on $\Sigma$ the vector $R^{\mu}$ becomes null-like
with respect to $G_{\mu\nu}^{-1},$ that is, $G_{\mu\nu}^{-1}R^{\mu}R^{\nu}=0,$
the signals propagate instantaneously (red propagation vectors from
cone B on Fig.~\ref{Cauchy 3d}) and one cannot guarantee the continuous
dependence on the initial data or even the existence and uniqueness
of the solution, see e.g. \cite{Vladimirov}. Using~(\ref{cs}) the
last inequality can be rewritten as \begin{equation}
c_{s}^{2}\left(1+\left(\dot{\phi}(\mathbf{x})\right)^{2}\frac{\mathcal{L}_{,XX}}{\mathcal{L}_{,X}}\right)>0.\label{joint condition}\end{equation}
Therefore, given Lagrangian $\mathcal{L}\left(\phi,X\right)$ and
hypersurface $\Sigma$ one has to restrict the initial data $\left(\phi(\mathbf{x}),\dot{\phi}(\mathbf{x})\right)$
by inequalities (\ref{hyper initial}) and (\ref{NO dips}) (or equivalently
(\ref{joint condition})), to have a well posed Cauchy problem. The
condition (\ref{joint condition}) is always satisfied in the subluminal
case for which $\mathcal{L}_{,XX}/\mathcal{L}_{,X}\geq0$. In addition,
we conclude that, if these conditions are satisfied everywhere on
the manifold $\mathcal{M}$ and the selected synchronous frame is
nonsingular in $\mathcal{M}$, then time $t$ plays the role of global
time and in accordance with the theorem about stable causality no
causal paradoxes arise in this case.

As a concrete application of the conditions derived let us find which
restrictions should satisfy the admissible initial conditions for
the low energy effective field theory with Lagrangian $\mathcal{L}(X)\simeq X-X^{2}/\mu^{4}+...$,
where $\mu$ is a cut off scale Ref.~\cite{Nima}. In this case (\ref{joint condition})
imply that not only $X\ll\mu^{4},$ but also $\left(\dot{\phi}(\mathbf{x})\right)^{2}\ll\mu^{4}$
and $\left(\vec{\nabla}\phi(\mathbf{x})\right)^{2}\ll\mu^{4}$. Note
that these restrictions can be rewritten in the Lorentz invariant
way: for example the first condition takes the form $\left(g^{\mu\nu}\partial_{\mu}t\partial_{\nu}\phi\right)^{2}\ll\mu^{4}.$

Finally let us note that even well-posed Cauchy problem cannot guarantee
the global existence of the unique solution for nonlinear system of
the equations of motion: for example, the solution can develop caustics
\cite{Kofman} or can become multi-valued \cite{Barbishov}.
\FIGURE{\label{Cauchy 3d}
\psfrag{T}[l]{$\partial_\mu t$} 
\psfrag{N}[l]{$N^\mu$}
\psfrag{S}[b]{\Large $\Sigma$}
\psfrag{r}[b]{$R^\mu$}
\psfrag{A}[b]{$A$} 
\psfrag{B}[b]{$B$}
\epsfig{file=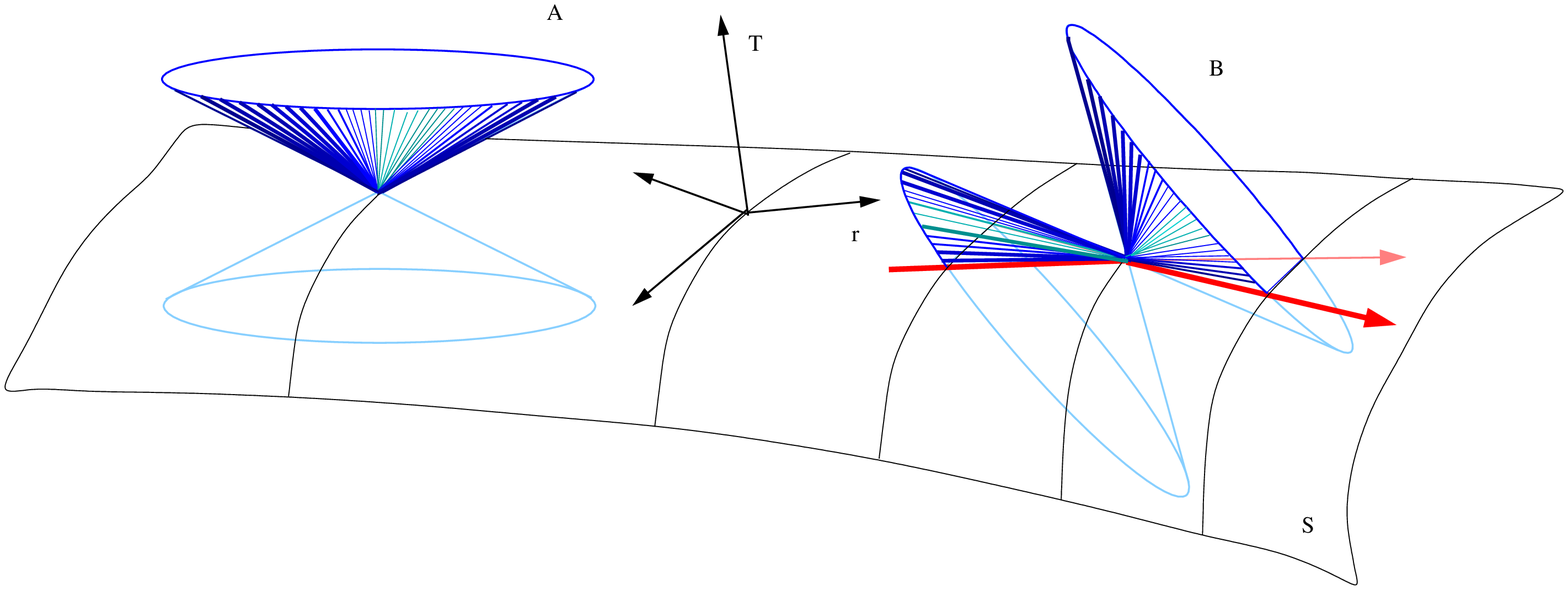,width=15cm,height=6.2cm}
\caption{The Cauchy problem for the equation of motion
of k-\emph{essence} is set up on the hypersurface $\Sigma$: $t=t_{0}$. 
The vector $R^{\mu}$ is tangential to $\Sigma$: $R^{\mu}\partial_{\mu}t=0$. 
For the hyperbolic equation of motion, the Cauchy problem is well
posed provided that $\partial_{\mu}t$ is timelike with respect to
$G^{\mu\nu}$ everywhere on $\Sigma$, or, equivalently, the hypersurface
$\Sigma$ is spacelike with respect to $G^{\mu\nu}$ (the cone $A$
in the figure). The cone $B$ represents an ill-posed Cauchy problem
for the hyperbolic equation. In particular the red propagation vectors
are tangent to $\Sigma$.}
}
\section{How to pose the initial conditions in a fast moving spacecraft?}
\label{SecV}
In this section we resolve \textquotedblleft paradoxes\textquotedblright\
which at first glance seems arising in the case of superluminal propagation
of perturbations \cite{Nima,Durrer2,Durrer} when one tries to formulate
the Cauchy problem in a fast moving spacecraft. To simplify the consideration
we restrict ourselves by purely kinetic k-\emph{essence}, for which
$\mathcal{L}\left(\phi,X\right)=\mathcal{L}(X)$ and assume that for
the background solution $X_{0}=const>0$ and $c_{s}>1$. This is a
reasonable approximation for more general backgrounds with $X_{0}>0$
on the scales much smaller than the curvature scale of the emergent
geometry $G_{\mu\nu}^{-1}$. There is always the preferred reference
frame $\left(t,x^{i}\right)$ in which the background is isotropic
and homogeneous. We refer to this frame as the \emph{rest frame}.
In the presence of an external source $\delta J$ equation (\ref{eompi})
in this frame takes the following form \begin{equation}
\partial_{t}^{2}\pi-c_{s}^{2}\bigtriangleup_{x}\pi=\xi\delta J,\label{EOM pi}\end{equation}
where $\xi\equiv\left(c_{s}^{2}/\mathcal{L}_{,X}\right)$, for details
see Appendix \ref{AppB}, equations (\ref{eq: Source rescalation}) and (\ref{Canonical}).
Now let us consider a spacecraft moving in $x$-direction with velocity
$v$ through the k-\emph{essence} background and denote the Lorentz
boosted comoving spacecraft coordinates by $\left(t^{\prime},x'^{i}\right)$.
As we have already mentioned above, if the velocity of the spacecraft
is larger than $c^{2}/c_{s}$ then the Cauchy problem for $\pi$ cannot
be well posed on the hypersurface $t^{\prime}=const$ %
\footnote{Throughout this section we explicitly write the speed of light $c$
and without loss of generality we assume $v>0$.%
}. After Lorentz transformation to comoving spacecraft frame, equation
(\ref{EOM pi}) becomes \begin{equation}
\left(1-\frac{v^{2}}{c^{2}}\right)^{-1}\left[\left(1-\frac{c_{s}^{2}v^{2}}{c^{4}}\right)\partial_{t^{\prime}}^{2}\pi-2v\left(1-\frac{c_{s}^{2}}{c^{2}}\right)\partial_{t^{\prime}}\partial_{x^{\prime}}\pi+\left(v^{2}-c_{s}^{2}\right)\partial_{x^{\prime}}^{2}\pi\right]-c_{s}^{2}\partial_{J}\partial_{J}\pi=\xi\delta J,\label{EOM normal Lorentz}\end{equation}
where prime denotes comoving coordinates and index $J=2,3,...$ stands
for the spatial directions other than $x^{\prime}$ {[}note that in
Ref.~\cite{Nima} the factor $\left(1-v^{2}/c^{2}\right)^{-1}$ in
front of squire brackets is missing{]}. For $v=c^{2}/c_{s}$ the second
time derivative drops out of (\ref{EOM normal Lorentz}) and the necessary
conditions for applicability of the Cauchy-Kowalewski theorem are
not satisfied; hence the existence and uniqueness of the solution
(\ref{EOM normal Lorentz}) are not guaranteed. For $v>c^{2}/c_{s}$
the necessary conditions of the Cauchy-Kowalewski theorem are met
and the unique solution of (\ref{EOM normal Lorentz}) exists; however,
this solution contains exponentially growing modes in the spatial
directions, perpendicular to $x^{\prime}$. Indeed, substituting \[
\pi\propto\exp\left(-i\omega't^{\prime}+ik_{x^{\prime}}x^{\prime}+ik_{J}x^{J}\right),\]
in (\ref{EOM normal Lorentz}) we find that in the boosted frame:
\begin{equation}
\omega'_{\pm}=\left(1-\frac{v^{2}c_{s}^{2}}{c^{4}}\right)^{-1}\left\{ k_{x^{\prime}}v\left(\frac{c_{s}^{2}}{c^{2}}-1\right)\pm c_{s}\sqrt{\left(1-\frac{v^{2}}{c^{2}}\right)\left[k_{x^{\prime}}^{2}\left(1-\frac{v^{2}}{c^{2}}\right)-\left(\frac{v^{2}c_{s}^{2}}{c^{4}}-1\right)k_{\bot}^{2}\right]}\right\} .\label{omega}\end{equation}
where we have denoted $k_{\bot}=\{ k_{J}\}$ and $k_{\bot}^{2}=k_{J}k_{J}$
. For $D=2$, when $k_{\bot}=0$, the frequencies $\omega$ are always
real and no instability modes exist (note that $v<c$). However, if
$D>2$ and $v>c^{2}/c_{s}$ then for \begin{equation}
k_{\bot}^{2}>k_{x^{\prime}}^{2}\left(\frac{1-v^{2}/c^{2}}{v^{2}c_{s}^{2}/c^{4}-1}\right),\label{forbidden K}\end{equation}
the general solution of (\ref{EOM normal Lorentz}) contains exponentially
growing modes. Note that these are the high frequency modes and hence
the instability would imply catastrophic consequences for the theory.
At first glance, this looks like a paradox, because equation (\ref{EOM pi}),
which has no unstable solutions in the rest frame, acquired exponentially
unstable solutions in the boosted frame. On the other hand, any solution
of~(\ref{EOM pi}) after performing the Lorentz transformation with
$v>c^{2}/c_{s}$ does not contain exponentially growing modes with
$k_{\bot}^{2}$ satisfying (\ref{forbidden K}). Indeed, given $\left(k_{x},k_{\bot}\right)$
in the rest frame one can perform the Lorentz transformation and obtain:\begin{equation}
\left\{ \omega^{\prime},\, k_{x^{\prime}},\, k'_{\bot}\right\} =\left\{ \frac{\omega+vk_{x}}{\sqrt{1-v^{2}/c^{2}}},\,\frac{k_{x}+\omega v/c^{2}}{\sqrt{1-v^{2}/c^{2}}},\, k_{\bot}\right\} ,\label{omegak}\end{equation}
were $\omega=\pm c_{s}\sqrt{k_{x}^{2}+k_{\bot}^{2}}$. Expressing
$\omega^{\prime}$ via $k_{x^{\prime}}$ and $k_{\bot}$ we again
arrive to (\ref{omega}). However, it follows from (\ref{omegak})
that if $v>c^{2}/c_{s}$ then the components of the Lorentz boosted
wavevector satisfy the condition\begin{equation}
k_{\bot}^{2}\leq k_{x^{\prime}}^{2}\left(\frac{1-v^{2}/c^{2}}{v^{2}c_{s}^{2}/c^{4}-1}\right),\label{allowed K}\end{equation}
and hence unstable modes are not present. This raises the question
whether the unstable modes which cannot be generated in the rest frame
of k-\emph{essence}, can nevertheless be exited by any physical device
in the spacecraft. We will show below that such device does not exist.
With this purpose we have to find first the Greens function in both
frames.

Let us begin with two-dimensional spacetime. In this case the retarded
Green's function for (\ref{EOM pi}) in the rest frame ($rf$) is
(see e.g. \cite{Vladimirov}):\begin{equation}
G_{R}^{\mathrm{rf}}(t,x)=\frac{1}{2c_{s}}\theta\left(c_{s}t-|x|\right).\label{Green rf}\end{equation}
In the boosted Lorentz frame it becomes\begin{equation}
G_{R}^{\mathrm{rf}}(t^{\prime},x^{\prime})=\frac{1}{2c_{s}}\theta\left(\frac{c_{s}\left(t^{\prime}+vx^{\prime}/c^{2}\right)-|x^{\prime}+vt^{\prime}|}{\sqrt{1-v^{2}/c^{2}}}\right).\label{Green rf boosted}\end{equation}
For $c_{s}v<c^{2}$, the Fourier transform of (\ref{Green rf boosted})
is the retarded in $t^{\prime}$ Green's function:\begin{equation}
G_{R}^{\mathrm{rf}}(t^{\prime},k^{\prime})=\frac{\theta(t^{\prime})}{2ic_{s}k^{\prime}}\,\left(e^{i\omega'_{+}t^{\prime}}-e^{i\omega'_{-}t^{\prime}}\right),\label{Fourier Green rf boosted slow}\end{equation}
whereas for $c_{s}v>c^{2}$ it is given by:\begin{equation}
G_{R}^{\mathrm{rf}}(t^{\prime},k^{\prime})=-\frac{\theta(t^{\prime})e^{i\omega'_{+}t^{\prime}}+\theta(-t^{\prime})e^{i\omega'_{-}t^{\prime}}}{2ic_{s}k^{\prime}}.\label{Fourier Green fr boosted fast}\end{equation}
This Green's function corresponds to the Feynman's boundary conditions
in the boosted frame. Thus, in the fast moving spacecraft, the \emph{retarded}
Green's function (\ref{Fourier Green fr boosted fast}), obtained
as a result of Lorentz transformation from (\ref{Green rf}) looks
like a mixture of the retarded {[}proportional to $\theta(t^{\prime})${]}
and the advanced {[}proportional to $\theta(-t^{\prime})${]} Green's
functions with respect to the spacecraft time $t^{\prime}$. In fact,
the situation is even more complicated. If from the very beginning
we work in the comoving spacecraft frame ($sc$), then solving~(\ref{EOM normal Lorentz})
we obtain the following expression for the retarded Green's function,
\begin{equation}
G_{R}^{\mathrm{sc}}(t^{\prime},k^{\prime})=\frac{\theta(t^{\prime})}{2ik^{\prime}c_{s}}\,\left(e^{i\omega'_{+}t^{\prime}}-e^{i\omega'_{-}t^{\prime}}\right).\label{rocket Green}\end{equation}
which coincides with equation (\ref{Green rf}), only if $c_{s}v<c^{2}$.
However, for fast moving spacecraft, $c_{s}v>c^{2}$, formula (\ref{rocket Green})
does not coincide with formula (\ref{Fourier Green fr boosted fast}).

The situation is more interesting in the four dimensional spacetime.
Similar to the 2d case, after we apply the Lorentz boost to the retarded
(in the rest frame) Green's function (see e.g. \cite{Vladimirov})\begin{equation}
G_{R}^{\text{rf}}(t,x^{i})=\frac{\theta\left(t\right)}{2c_{s}\pi}\delta\left(c_{s}^{2}t^{2}-|x|^{2}\right),\label{4d Green rest}\end{equation}
and calculate its Fourier transform (see Appendix \ref{AppE} for the details)
we find that for the slowly moving spacecraft, $vc_{s}<c^{2}$, \begin{equation}
G_{R}^{\text{rf}}(t^{\prime},k^{\prime})=\frac{\theta\left(t^{\prime}\right)}{2ic_{s}}\left(k_{x^{\prime}}^{2}+k_{\bot}^{2}\frac{1-c_{s}^{2}v^{2}/c^{4}}{1-v^{2}/c^{2}}\right)^{-1/2}\left(\text{e}^{i\omega'_{+}t^{\prime}}-\text{e}^{i\omega'_{-}t^{\prime}}\right).\label{Retarded green on slow rocket}\end{equation}
That is, the resulting Green's function is also retarded with respect
to the spacecraft time $t^{\prime}$. On the other hand, for the fast
moving spacecraft, $vc_{s}>c^{2}$, we obtain:\begin{equation}
G_{R}^{\mathrm{rf}}(t^{\prime},k^{\prime})=-\frac{1}{2ic_{s}}\left(k_{x^{\prime}}^{2}+k_{\bot}^{2}\frac{1-c_{s}^{2}v^{2}/c^{4}}{1-v^{2}/c^{2}}\right)^{-1/2}\left(\theta\left(t^{\prime}\right)\text{e}^{i\omega'_{+}t^{\prime}}+\theta\left(-t^{\prime}\right)\text{e}^{i\omega'_{-}t^{\prime}}\right).\label{Fast moving green correct}
\end{equation}
Similar to the 2d case formula (\ref{Fast moving green correct})
is the Feynman Green's function in the spacecraft frame. Note that
formula (\ref{Fast moving green correct}) can be rewritten as: \begin{eqnarray}
G_{R}^{\mathrm{rf}}(t^{\prime},k^{\prime}) & = & \frac{1}{2c_{s}}\left(k_{x^{\prime}}^{2}+k_{\bot}^{2}\frac{1-c_{s}^{2}v^{2}/c^{4}}{1-v^{2}/c^{2}}\right)^{-1/2}\times\label{fm}\\
 & \times & \exp\left(-ik_{x^{\prime}}vt^{\prime}\frac{1-c_{s}^{2}/c^{2}}{1-c_{s}^{2}v^{2}/c^{4}}-\frac{1-v^{2}/c^{2}}{c_{s}^{2}v^{2}/c^{4}-1}c_{s}\left|t^{\prime}\right|\sqrt{k_{\bot}^{2}\frac{1-v^{2}/c^{2}}{c_{s}^{2}v^{2}/c^{4}-1}-k_{x^{\prime}}^{2}}\right).\notag\end{eqnarray}
 It is obvious from here that the modes with large $k_{\bot}$are
exponentially suppressed and therefore very high frequency source
$\delta J$ cannot excite perturbations with $k_{\bot}^{2}$ satisfying
inequality (\ref{forbidden K}).

In the spacecraft frame the retarded Green's function calculated directly
for Fourier modes of (\ref{EOM normal Lorentz}) is:\[
G_{R}^{\text{sc}}(t^{\prime},k^{\prime})=\frac{\theta\left(t^{\prime}\right)}{2ic_{s}}\left(k_{x^{\prime}}^{2}+k_{\bot}^{2}\frac{1-c_{s}^{2}v^{2}/c^{4}}{1-v^{2}/c^{2}}\right)^{-1/2}\left(\text{e}^{i\omega'_{+}t^{\prime}}-\text{e}^{i\omega'_{-}t^{\prime}}\right).\]
It coincides with Green's function (\ref{Retarded green on slow rocket}),
obtained by applying the Lorentz transformation, only in the case
of slow motion with $v<c^{2}/c_{s}$. However, the results drastically
differ for the fast moving spacecraft - compare equations (\ref{Retarded green on slow rocket})
and (\ref{Fast moving green correct}). The function $G_{R}^{\text{sc}}(t^{\prime},k^{\prime})$
contains exponentially growing modes for sufficiently large $k_{\bot}$
and it's Fourier transform to coordinate space $G_{R}^{\mathrm{sc}}(t^{\prime},x^{\prime})$
does not exist. Physically this means that we have failed to find
the Green's function, which describes the propagation of the signal
which the source $\delta J$ in the fast moving spacecraft tries to
send in the direction of growing $t^{\prime}$. Instead, the response
to any source in the spacecraft is always driven by (\ref{fm}) (or
the Lorentz transformed Green's function in the rest frame (\ref{4d Green rest})).
Because we cannot send a signal in the direction of growing $t^{\prime}$
one cannot associate growing $t^{\prime}$ with the arrow of time
contrary to the claims in \cite{Durrer2}.

Now we will discuss in more details how the problem of initial conditions
for perturbations $\pi$ must be correctly formulated in the fast
moving spacecraft. The first question here whether the fast moving
astronaut can create an arbitrary initial field configurations $\pi$
and $\dot{\pi}$ at a given moment of his proper time $t_{1}^{\prime}=const.$
This hypersurface is not \emph{space-like} with respect to the metric
$G_{\mu\nu}^{-1}$ and therefore as it follows from the consideration
in the previous section the Cauchy problem is not well posed on it.
Hence not all possible configurations are admissible on this hypersurface
but only those which could be obtained as a result of evolution of
some initial configuration chosen on the hypersurface which is simultaneously
spacelike with respect to both metrics $g_{\mu\nu}$ and $G_{\mu\nu}^{-1}.$
If the astronaut disturbs the background with some device (source
function $\delta J$) which he/she switches off at the moment of time
$t_{1}^{\prime}$, then the resulting configuration of the field on
the hypersurface $t_{1}^{\prime}=const$ obtained using the correct
Green's function (\ref{fm}) will always satisfy the conditions needed
for unambiguous prediction of the field configuration everywhere in
the spacetime irrespective of the source $\delta J(x).$ The presence
of the advanced mode in this Green's function plays an important role
in obtaining a consistent field configuration on $t_{1}^{\prime}=const.$
Thus we see that not \textquotedblleft everything\textquotedblright\ is
in the hand of the astronaut: he has no \textquotedblleft complete
freedom\textquotedblright\
in the choice of the \textquotedblleft initial\textquotedblright\ field
configuration at time $t_{1}^{\prime}.$ Nonrecognition of this fact
leads to the fictitious causal paradoxes discussed in the literature
\cite{Durrer,Durrer2}.

For a slowly moving spacecraft, $v<c^{2}/c_{s}$, the retarded Green's
function in the rest frame is transformed in the retarded Green's
function in the spacecraft frame. Therefore we can obtain any a priori
given field configuration on the hypersurface $t_{1}^{\prime}=const$
by arranging the source function $\delta J$ in the corresponding
way. Thus, the choice of the initial conditions for the perturbations
at $t_{1}^{\prime}=const$ is entirely in the hand of the astronaut.
This is in complete agreement with our previous consideration because
in the slowly moving spacecraft the hypersurface $t_{1}^{\prime}=const$
is spacelike with respect to both metrics.

The appearance of the advance part in the correct Green's function
for the fast moving spacecraft still looks a little bit strange because
according to the clocks of the astronaut the head of the spacecraft
can \textquotedblleft
feel\textquotedblright\ signals sent at the same moment of time
by a device installed on the stern of the spacecraft. However, in
this case the proper time of the astronaut is simply not a good coordinate
for the time ordering of the events at different points of the space
related by the k-\emph{essence} superluminal signals. The causality
is also preserved in this case but it is determined by the superluminal
k-\emph{essence} cone which is larger than the light cone and as we
have already seen no causal paradoxes arise in this case. If the astronaut
synchronizes his clocks using the superluminal sonic signals then
the new time coordinate $\tilde{t}$ becomes a good coordinate for
the time ordering of the causal events in different points of the
space. The hypersurface $\tilde{t}=const$ being spacelike in both
metrics can then be used as the initial hypersurface for the well
posed Cauchy problem in the fast moving spacecraft, that is, any initial
configuration of the field can be freely created by the astronaut
on this hypersurface. In the \textquotedblleft well synchronized'
reference frame $\left(\tilde{t},\tilde{x},\tilde{y},\tilde{z}\right)$
the equation of motion for perturbations (\ref{EOM pi}) takes the
same form as in the rest frame of the k-\emph{essence} background:
\begin{equation}
\partial_{\tilde{t}}^{2}\pi-c_{s}^{2}\bigtriangleup_{\tilde{x}}\pi=\xi\delta J.\label{EOM cs Lorentz}\end{equation}
It follows from here that\[
\omega_{\pm}=\pm c_{s}\sqrt{k_{\tilde{x}}^{2}+k_{\tilde{y}}^{2}+k_{\tilde{z}}^{2}},\]
and hence no exponentially growing modes exist for any $k_{\tilde{x}},$
$k_{\tilde{y}}$ and $k_{\tilde{z}}$.

The causal Green's function in the spacecraft frame contains only
the retarded with respect to the time $\tilde{t}$ part. For example,
in four-dimensional spacetime it is given by \begin{equation}
G_{R}^{\mathrm{sc}}(\tilde{t},\tilde{x}^{i})=\frac{\theta\left(\tilde{t}\right)}{2c_{s}\pi}\delta\left(c_{s}^{2}\tilde{t}^{2}-|\tilde{x}^{i}|^{2}\right).\label{Green rocket correct}\end{equation}
This result can be obtained either by applying the Lorentz transformation
with the invariant speed $c_{s}$ to (\ref{Green rf}), or directly
by solving~equation (\ref{EOM cs Lorentz}). Thus, no paradoxes with
Green's functions arise for the superluminal perturbations. The same
conclusions are valid in 4d spacetime.
\FIGURE{\label{Figure Fluid}
\psfrag{T}[l]{\large $t$} \psfrag{X}[b]{\large $x$} \psfrag{O}[l]{\large $O$} \psfrag{TT}[l]{\large $t'$} \psfrag{XX}[b]{\large $x'$} \psfrag{XXX}[b]{\large $\tilde{x}$} \psfrag{light}[t][r]{\textit{light cone}} \psfrag{acoustic cone}[b][]{\textit{acoustic cone}}\psfrag{Y}[]{\large $x_{J}$}
\epsfig{file=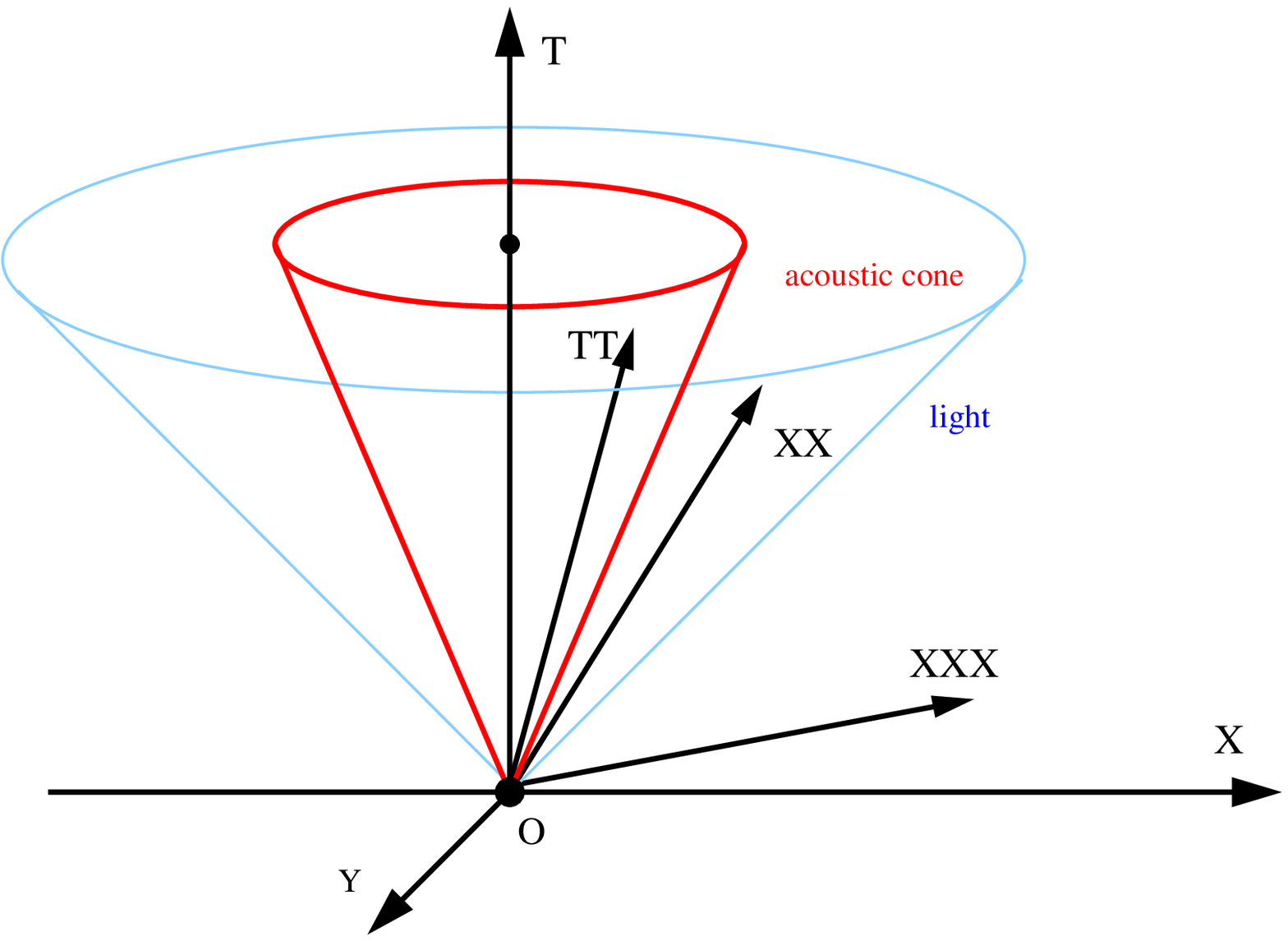,width=11cm,height=7cm}
\caption{It is shown how one can create a would -be
``paradox'' similar to that discussed in this section, without involving
any superluminal signals. The fluid is at rest and the perturbations
propagate subluminally in the fluid, $c_{s}<c$. The reference frame
$(t',\, x')$ is connected to the rest frame by the Lorentz boost
with the invariant speed $c_{s}$. If the boost speed $v$ is such
that $c_{s}/c<v/c_{s}<1$, then the hypersurface of constant $t'$
is inside the light cone and the Cauchy problem for the electromagnetic
field is ill posed in this reference frame. Instead, one should use
the ``correct'' frame $(\tilde{t},\,\tilde{x})$, obtained by the
Lorentz boost with the invariant fastest speed $c=1$. In this frame
the Cauchy problem is well-posed.}
}

To make the consideration above even more transparent we conclude
this section by considering analogous situation with \emph{no} superluminal
signals involved. Namely, we take a fluid at rest with a \emph{subluminal}
speed of sound, $c_{s}<c$. Then we can make the Lorentz transformation
using the invariant speed $c_{s}$: \[
t^{\prime}=\frac{t-vx/c_{s}^{2}}{\sqrt{1-v^{2}/c_{s}^{2}}},\,\,\, x^{\prime}=\frac{x-vt}{\sqrt{1-v^{2}/c_{s}^{2}}},\,\,\, x_{J}^{\prime}=x_{J}.\]
If the speed $v$ is such that $c_{s}/c<v/c_{s}<1$, then the hypersurface
of constant $t^{\prime}$ is \emph{inside} the light cone (see Fig.~\ref{Figure Fluid})
and it is obvious that one cannot formulate the Cauchy problem for
the electromagnetic field on the hypersurface $t^{\prime}=const$.
Instead, the Cauchy problem for the electromagnetic field can be well
posed on the hypersurface $\tilde{t}=const$ defined by the \textquotedblleft
correct\textquotedblright\ Lorentz transformation, with the invariant
speed $c$: \[
\tilde{t}=\frac{t-vx/c^{2}}{\sqrt{1-v^{2}/c^{2}}},\,\,\,\tilde{x}=\frac{x-vt}{\sqrt{1-v^{2}/c^{2}}},\,\,\,\tilde{x}_{J}=x_{J},\]
(see Fig.~\ref{Figure Fluid}). This consideration is fully equivalent
to those one above with the only replacement $c_{s}\leftrightarrow c$.

Thus we have shown that no physical paradoxes arise in the case when
we have superluminal propagation of small perturbations on the background.
\section{Chronology protection}
\label{SecVI}
It was claimed in \cite{Nima} that the theories with superluminal
propagation are plagued by closed causal curves (CCC). We will argue
here that the superluminal propagation cannot be the sole reason for
the appearance of CCC and moreover this problem can be avoided in
this case in the same way as in General Relativity.

It is well know that General Relativity admits the spacetimes with
the closed causal curves without involving any superluminal fields
into consideration. Among examples of such spacetimes are: Gödel's
cosmological model \cite{Godel}, Stockum's rotating dust cylinder
\cite{Stockum}, wormholes \cite{wormhole}, Gott's solution for two
infinitely long strings \cite{Gott} and others \cite{otherCCC}.
A prominent time-machine model was suggested recently by Ori \cite{Ori}.
In this model, made solely of vacuum and dust, the spacetime evolves
from a regular normal asymptotically flat state without CCCs and only
later on develops CCCs without violating the weak, dominant and strong
energy conditions. Thus, we see that initially \textquotedblleft good\textquotedblright\ spacetime
might in principle evolve to a state where the chronology is violated
and the General Relativity does not by itself explains these strange
phenomena. Therefore one needs to invoke some additional principle(s)
to avoid the pathological situations with CCCs. With this purpose
Hawking suggested the \textit{Chronology Protection Conjecture}, which
states that the laws of physics must prohibit the appearance of the
closed timelike curves \cite{Chronology}. In \cite{Chronology} it
was argued that in the situation when the timelike curve is ready
to close, the vacuum polarization effects become very large and the
backreaction of quantum fields prevents the appearance of closed timelike
curves.

Similarly to General Relativity, one might assume that in the case
of superluminal propagation the chronology protection conjecture is
valid as well. For example, the chronology protection was already
invoked to exclude the causality violation in the case of two pairs
of Casimir plates \cite{Liberati}, in which photons propagate faster
than light due to the Scharnhorst effect \cite{plates}.

Once we employ the chronology protection principle, no constructions
admitting CCCs, similar to those presented in \cite{Nima}, may become
possible.

In fact, the first example in \cite{Nima} with two finite fast moving
bubbles made of superluminal scalar field (see Fig.~2 in Ref.~\cite{Nima})
is quite similar to the \textquotedblleft time machine\textquotedblright\
involving two pair of Casimir plates. In the latter case the chronology
protection excludes the existence of CCCs. Here the situation is a
little bit more involved. In the example with the bubbles the background
is not a free solution of the equation of motion (\ref{EOM}). Indeed
as it was pointed out in \cite{Nima} the fast moving bubbles have
to be separated in the direction orthogonal to the direction of motion.
On the other hand they have to be connected by light. However, if
this were a free solution, then the bubbles would expand with the
speed of light and collide at the same moment of time, or even before
the closed causal curve would be formed. Thus an external source $J(x)$
of the scalar field is required in order to produce this acausal background.
However, without clear idea about the origin of this source and possible
backreaction effects the physical interpretation of this \textquotedblleft time
machine\textquotedblright\
is obscure. It is well known that admitting all possible sources of
gravitational field one can obtain almost any possible even acausal
solutions in general relativity. Finally, generalizing the Hawking
conjecture to the case of scalar field one can argue that the backreaction
of quantum fluctuations of perturbations $\pi$ around $\phi$ become
large before CCC is formed thus destroying the classical solution
imposed by the external source $J(x)$ and preventing the formation
of CCC.

The other example considered in \cite{Nima} involves non-linear electrodynamics.
The electromagnetic field is created by charge currents, serving as
a source. Thus, unlike the previous example, one can control the strength
of the field, simply changing the configuration of charges. The electromagnetic
part of the Lagrangian is the \textquotedblleft
wrong\textquotedblright -signed Euler-Heisenberg Lagrangian: \begin{equation}
\mathcal{L}=-\frac{1}{4}F_{\mu\nu}F^{\mu\nu}-\alpha\left(F_{\mu\nu}F^{\mu\nu}\right)^{2}+...,\label{NLEM}\end{equation}
with a small positive $\alpha$. For such a system the propagation
of light in a non-trivial background is superluminal. As a consequence,
a cylindrical capacitor with the current-carrying solenoid leads to
the appearance of the CCCs, provided the electrical and magnetic fields
inside the capacitor are large enough (see Fig.~3 in \cite{Nima}).
In this case one may invoke a simplified version of the chronology
protection conjecture. In fact, let us begin with some \textquotedblleft good\textquotedblright\ initial
conditions in the capacitor, namely, with electric and magnetic fields
being not too large, so that no CCCs exist. Then we increase the current
in the solenoid and the voltage between the plates of a capacitor
in order to increase the strength of the fields. When the causal curves
become almost closed, the expectation value of the energy-momentum
tensor for the \textquotedblleft quasi-photons\textquotedblright\ on
this classical electromagnetic background becomes very large due to
the quantum vacuum polarization effects. In the limit when the causal
cone becomes horizontal, the energy density of the field in the capacitor
tends to infinity and the capacitor will be broken before the CCCs
will be formed.

Thus we conclude that concerning the causal paradoxes the situation
in the theories with superluminal propagation on the non-trivial backgrounds
is not much worse than in General Relativity. In fact, in this respect
the similarity between these two theories goes even much deeper than
it looks at the first glance. For example, let us imagine a time machine
which is constructed with the help of superluminal propagation in
non-trivial background produced by the external source $J\left(x\right)$,
e.g., similar to those described in \cite{Nima}. Then we can identify
the \emph{effective metric} $G^{\mu\nu}$ for this system with the
\emph{gravitational metric} $g^{\mu\nu}$ of some spacetime produced
by an energy-momentum tensor $T_{\mu\nu}^{(J)}\left(x\right)$. Put
differently, once having the effective metric, we can find spacetime
where the gravitational metric is $G^{\mu\nu}$. In this spacetime
the time machine exists as well. Remarkably, now the gravitation (or
light) signals are used to make CCCs. The spacetime with the metric
$G^{\mu\nu}$ is the solution of Einstein equations with the energy-momentum
tensor $T_{\mu\nu}^{(J)}$ calculated substituting the metric in the
Einstein equations$.$ After that one could try to find such theories
and such fields configurations on which their resulting energy momentum
tensor is equal to $T_{\mu\nu}^{(J)}\left(x\right)$ consistently
with equations of motion. One can, in principle, argue, that in the
case when the CCCs exist the energy-momentum tensor mights have some
undesired properties, for example, it would violate the Week Energy
Condition (WEC). However, in several known examples with CCCs the
WEC is satisfied, see, e.g. Refs.~\cite{Godel,Gott,Stockum}. Moreover,
the system found by A.~Ori \cite{Ori} possesses CCCs and satisfies
the week, dominant and strong energy conditions. Thus the violation
of the energy conditions is not an inherent property of the spacetimes
with CCCs. Therefore the question, whether the spacetime constructed
by the procedure described above requires \textquotedblleft bad\textquotedblright\ energy-momentum
tensor or not, must be studied separately in each particular case.

Moreover the correspondence $G^{\mu\nu}\leftrightarrow g^{\mu\nu}$,
$J\leftrightarrow T_{\mu\nu}^{(J)}$ can also be used to learn more
about \emph{Chronology Protection Conjecture} and time-machines in
General Relativity with the help of more simple theory. It is well-known
that Analogue Gravity \cite{Analog} gives more simple and intuitively
clear way to investigate the properties of Hawking radiation, the
effects of Lorentz symmetry breaking, transplanckian problem \emph{etc.,}
by using the small perturbations in the fluids instead of direct implication
of General Relativity. In a similar way, \emph{analogue time-machine}
or analogue \emph{Chronology Protection Conjecture} may provide one
with a tool to check \emph{Chronology Projection Conjecture} and the
possibility of construction of time machines in General Relativity.
\section{Is the gravitational metric universal?}
\label{SecVII}
It was argued in~\cite{Ellis} that the same causal limits apply
to all fields independent of the matter present, thus endowing the
gravitational metric with the universal role. However, in the theories
under consideration the causal limit is governed not by the gravitational
metric $g_{\mu\nu},$ but by the effective acoustic metric $G_{\mu\nu}^{-1}$
and hence the gravity loses its universal role in this sense. Nevertheless,
here we argue, that even in the case of the spontaneously broken Lorentz
invariance with a superluminal propagation the gravitational metric
$g_{\mu\nu}$ still keeps its universal role in the following sense.
First in accordance with the discussion in \ref{SecIV}
we remind that the Cauchy hypersurface for the field $\phi$ should
anyway be a spacelike one in the gravitational metric. Thus in order
to produce a background which breaks the Lorentz invariance one has
to respect the usual causality governed by the gravitational metric
$g_{\mu\nu}$. Moreover, if a clump of the scalar field is created
in a finite region surrounded by a trivial background, then the boundaries
of the clump will generically propagate with the speed of light.

Indeed, let us consider a finite lump of non-trivial field configuration
with smooth boundaries (see Fig.~\ref{universality}) and assume
that the initial data $\left(\phi(\mathbf{x}),\dot{\phi}(\mathbf{x})\right)$
are specified in some finite spatial region $R$. These initial data
are smooth everywhere (see Fig.~\ref{universality}) and satisfy
the conditions (\ref{joint condition}) and (\ref{hyper initial}),
in particular the first derivatives of the field $\phi$ are continuous
everywhere including the boundaries of the clump. If the system described
by action (\ref{action}) has at least one trivial solution $\phi=\phi_{\text{triv}}=const$
with non-pathological acoustic geometry, then, as it follows from
(\ref{EOM}) and (\ref{Gdown}), the Lagrangian $\mathcal{L}(\phi,X)$
is at least twice differentiable at $(\phi,X)=(\phi_{\text{triv}},0)$
and moreover $\mathcal{L}_{,X}(\phi_{\text{triv}},0)\neq0$. Thus
for the theories of this type we have \begin{equation}
\mathcal{L}(\phi,X)\simeq V(\phi)+K_{1}(\phi)X+K_{2}(\phi)X^{2}+...\label{11}\end{equation}
in the vicinity of the trivial solution $\phi_{\text{triv}}$ %
\footnote{In particular, it was required in \cite{kdefect} that for models
allowing topological k-\emph{defects}, the asymptotic behavior near
the trivial vacuum $X=0$ (at the spatial infinity) is of the form
(\ref{11})%
}. And as expected we conclude that the speed of sound for the small
perturbations is equal to the speed of light in the vicinity of $\phi_{\text{triv}}$
because any trivial solution and in particular a possible vacuum solution
$\phi=0$ does not violate the Lorentz invariance. Moreover, close
to the boundaries of the clamp the initial data $\left(\phi(\mathbf{x}),\dot{\phi}(\mathbf{x})\right)$
can be considered as small perturbation around the trivial background
and therefore the front of the clump propagates exactly with the speed
of light in the vacuum. Thus, without preexisting nontrivial configuration
of the scalar field the maximum speed of propagation never exceeds
the speed of light and the causality is entirely determined by the
usual gravitational metric only.
\FIGURE{\label{universality}
\psfrag{R}[b]{$R$} 
\psfrag{S}[b]{\Large $\Sigma$}
\psfrag{F1}[b][l]{\Large $\phi(\bf{x})$}
\psfrag{F2}[b][l]{\Large $\dot{\phi}(\bf{x})$}
\epsfig{file=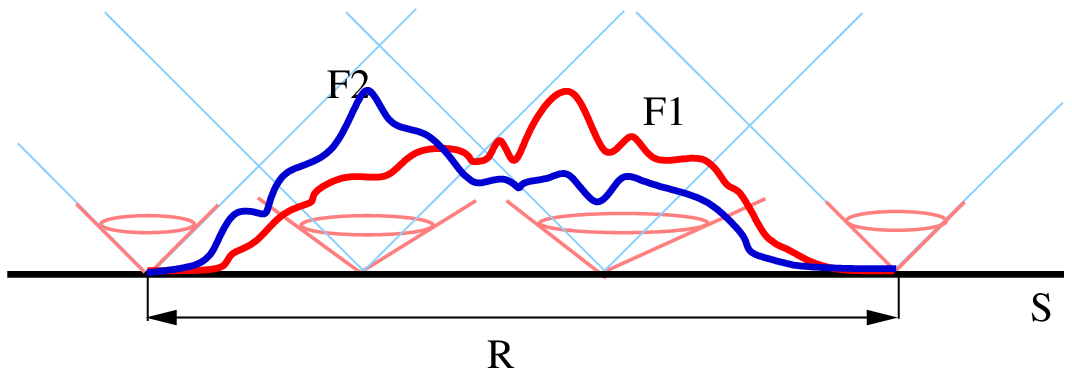,width=15cm,height=5cm}
\caption{ The figure shows that the gravitational metric
$g_{\mu\nu}$ keeps its universal meaning even if the small perturbations
on the non-trivial backgrounds propagate superluminally. If in the
initial moment of time the non-trivial configuration of the field
$\phi$ is localized in the finite region $R$ on a spacelike in $g_{\mu\nu}$
hypersurface $\Sigma$, and beyond this region the field $\phi$ is
in its vacuum state $\phi=const$, then the front of the solution
always propagates with the speed of light. The blue lines correspond
to the light rays. The pink cones represent the influence cones for
k-\emph{essence}. On the boundary of $R$ the influence cones are
equal to light cones. }
}

If we abandon the condition of the regularity of the emergent geometry
$G_{\mu\nu}^{-1}$, but still require that the Lagrangian is analytic
function of $X$ in the neighborhood of $X=0$, then the speed of
propagation in vacuum is always \emph{smaller} than the speed of light.
Indeed in this case the speed of sound $c_{s}$ is:\[
c_{s}^{2}=\frac{1}{\left(1+2\left(n-1\right)\right)}<1,\]
where $n$ is the power of the first non-zero kinetic term in (\ref{11}).

To demonstrate explicitly the points stated above we will find now
exact solitonic solutions in the purely kinetic k-\emph{essence} theories
with Lagrangian $\mathcal{L}\left(X\right)$ and verify that these
solitons propagate in the Minkowski spacetime with the speed of light.
Assuming that the scalar field depends only on $\theta\equiv x+vt$
and substituting $\phi=\varphi\left(\theta\right)$ in equation (\ref{EOM})
we find that this equation reduces to \begin{equation}
\mathcal{L}_{,X}\varphi_{,\theta\theta}\left(v^{2}-1\right)+\mathcal{L}_{,XX}\varphi_{,\theta\theta}\varphi_{,\theta}^{2}\left(v^{2}-1\right)^{2}=0,\label{1}\end{equation}
This equation is trivially satisfied for $v=\pm1$, that is, there
exist solitary waves $\varphi\left(x\pm t\right)$ propagating with
the speed of light. They are solutions corresponding to rather special
initial conditions $\phi_{0}\left(x\right)=\varphi\left(x\right)$
and $\dot{\phi}_{0}\left(x\right)=\pm\varphi\left(x\right)$. Note
that the general solutions are not a superposition of these solitonic
solutions because the equation of motion is nonlinear. Assuming that
$v\neq\pm1$ we find that (\ref{1}) is satisfied by either nonlocalized
solution $\phi=x\pm vt+const,$ or it reduces to:\begin{equation}
\mathcal{L}_{,X}+\mathcal{L}_{,XX}\varphi_{,\theta}^{2}\left(v^{2}-1\right)=\mathcal{L}_{,X}+2X\mathcal{L}_{,XX}=0,\label{2}\end{equation}
This algebraic equation is trivially satisfied for all $X$ if $\mathcal{L=}f\left(\phi\right)\sqrt{X}-V\left(\phi\right).$
This is the case when the perturbations propagate with the infinite
speed on any background \cite{Cuscuton}. For more general Lagrangians
$\mathcal{L}\left(X\right)$ equation (\ref{2}) can be solved algebraically
to obtain a particular $X_{0}=\frac{1}{2}\varphi_{,\theta}^{2}\left(v^{2}-1\right)=const$.
The only solutions of this last equation are either $\varphi\left(x\pm t\right)$
or trivial solutions $\phi_{\text{triv}}=const$. For the \ Born-Infeld
Lagrangian \cite{Born-Infel} the exact solutions of this type were
found in \cite{Barbishov} (see also \cite{Nonlinear}). For more
complicated Lagrangian, for example, of the form $\mathcal{L}=\mathcal{K}(X)+V(\phi)$
there exist solitonic solutions with $v<1$ \cite{kdefect}.

Thus, we have shown that under reasonable restrictions on the theory
the field configurations localized in trivial vacuum never propagate
faster than light. Therefore the causal limit for these localized
configurations is always governed by the usual gravitational metric.
\section{Discussion}
\label{SecVIII}
In this paper we have considered the k-\emph{essence}-like scalar
fields with the Lorentz invariant action (\ref{action}) and have
studied the issues of causality and Cauchy problem for such theories.
These questions are non-trivial because small perturbations $\pi$
on backgrounds $\phi_{0}$ can propagate faster-than-light. The perturbations
\textquotedblleft
feel\textquotedblright\ the effective metric, $G^{\mu\nu}$ given
by~(\ref{Gdown}), which is different from the gravitational metric
$g^{\mu\nu}$, if the Lagrangian $\mathcal{L}$ is a non-linear function
of $X$ and the background is nontrivial $\partial_{\mu}\phi_{0}\neq0$.
We have derived the action for the perturbations on an arbitrary background
and have shown that these perturbations \textquotedblleft feel\textquotedblright\ the
emergent geometry $G_{\mu\nu}^{-1}$. The influence cone determined
by $G_{\mu\nu}^{-1}$ is larger than those one determined by metric
$g_{\mu\nu}$ provided $\mathcal{L}_{,XX}/\mathcal{L}_{,X}<0$ \cite{Susskind,Rendall,Halo}.
Thus perturbations can propagate with the speed exceeding the speed
of light. In this case the background serves as a new \emph{aether}
and preselects the preferred reference frame. This is why the causal
paradoxes arising in the presence of \emph{tachyons} %
\footnote{Do not confuse them with field theoretical tachyons with $m^{2}<0$.%
} (superluminal particles in the Minkowski vacuum) do not appear here\emph{.}
In particular, we have shown that in physically interesting situations,
namely, cosmological solutions and for the case of a black hole surrounded
by an accreting fluid, the closed timelike curves are absent and hence
we cannot send the signal to our own past using the superluminal signals
build out of the \textquotedblleft superluminal\textquotedblright\ scalar
field perturbations. Thus, the k-\emph{essence} models, which generically
possess the superluminal propagation, \emph{do not lead to the causal}
\emph{paradoxes}, contrary to the claim in~\cite{Durrer,Durrer2}.

We have shown how to pose correctly the Cauchy problem for the k-\emph{essence}
fields with superluminal propagation, which sometimes might seem problematic
\cite{Nima}. The correct initial Cauchy hypersurface $\Sigma$ must
simultaneously be spacelike with respect to both gravitational metric
$g_{\mu\nu}$ and the effective metric $G_{\mu\nu}^{-1}.$ Because
the effective metric $G_{\mu\nu}^{-1}$ itself depends on the values
of the field $\phi$ and its first derivatives, the initial value
problem must be set up in a self-consistent manner: in addition to
the usually assumed hyperbolicity condition (\ref{HYPER}), one must
require that the field $\phi$ and its derivative on $\Sigma$ must
satisfy the inequality (\ref{NO dips}). In particular, in the case
of spacecraft which has very large velocity with respect to the homogeneous
background of the k-\emph{essence}, the latter conditions are violated
on the hypersurface of constant astronaut proper time. Therefore no
physical devices are able to produce an arbitrary configuration of
perturbations on this hyperspace.

It was found in \cite{Nima} that in the theories under consideration
one can have the backgrounds possessing the closed causal curves (CCCs).
However, as we have argued above, this is not directly related to
the superluminal propagation. In fact, the situation here is very
similar to the situation in General Relativity, where one can also
have the manifolds with the closed causal curves although the speed
of propagation is always limited by the speed of light. In this respect
the situation in the theories with the superluminal propagation is
not worse than in General Relativity. To avoid causal paradoxes in
General Relativity, Hawking suggested the \emph{Chronology Protection
Conjecture}, which states that the quantum effects and, in particular,
vacuum polarization effects can prevent the formation of the closed
timelike curves \cite{Chronology}. Similarly to Hawking one may argue
that in the case of the superluminal propagation the \emph{Chronology
Protection Conjecture} can be valid as well. In fact, this conjecture
was already invoked to exclude the causality violation in the case
of two pairs of Casimir plates \cite{Liberati}. Once we employ the
\emph{Chronology Protection Conjecture}, no constructions admitting
CCCs, similar to those presented in \cite{Nima}, are possible.

Sometimes the {}``superluminal'' theories are criticized in the
literature on the basis of general, or better to say, aesthetic grounds.
For example, Ref.~\cite{Ellis} claims: {}``The spacetime metric
is preferred in terms of clock measurements and free fall (geodesic)
motion (including light rays), thus underlying General Relativity's
central theme of gravity being encoded in spacetime curvature.''
Although this argument is not more than the matter of taste, we would
rather prefer to have General Relativity as a theory which keeps its
(restricted) universal meaning even in the presence of superluminal
propagation. We have argued that under physically reasonable assumptions
and without a preexisting nontrivial background the causality is governed
by metric $g_{\mu\nu}$. Indeed, if initially the field $\phi$ is
localized within some finite region of space surrounded by vacuum,
then the border of this region propagates with the speed of light
and it is impossible to send signals faster than light.

\acknowledgments
We are very thankful to Camille Bonvin, Chiara Caprini, Sergei Dubovsky,
Ruth Durrer, Valery Frolov, Robert Helling, Mattew Kleban, Lev Kofman,
Stefano Liberati, Alan Rendall, Sergei Sibiryakov, Ilya Shapiro, Alexey
Starobinsky, Leonard Susskind, Matt Visser, and especially Sergei
Winitzki for very useful discussions. It is a pleasure to thank Sergei
Winitzki for helpful comments on the first version of the manuscript.
E.B. thanks Alexander von Humboldt foundation and INFN for support.
A.V. would like to thank the theory group of Laboratori Nazionali
del Gran Sasso, INFN, and organizers and staff of Les Houches Summer
School for hospitality during the preparation of this manuscript and
during earlier stages of this project respectively.

\appendix
\section{Characteristics and superluminal propagation}
\label{AppA}
Let us consider scalar field $\phi$ interacting with external source
$J\left(x\right)$. The equation of motion for the scalar field is
\begin{equation}
\tilde{G}^{\mu\nu}\nabla_{\mu}\nabla_{\nu}\phi+\varepsilon_{,\phi}=J\label{EOM with Source}\end{equation}
where metric $\tilde{G}^{\mu\nu}$ is given by (\ref{Metric}) and
for brevity we use the \textquotedblleft hydrodynamic\textquotedblright\
notation $\varepsilon\left(X,\phi\right)=2X\mathcal{L}_{,X}-\mathcal{L}$
(see Appendix \ref{AppD}). Suppose $\phi_{0}$ is the background solution of
(\ref{EOM with Source}) in the presence of source $J_{0}\left(x\right)$
and gravitational metric $g_{\mu\nu}\left(x\right)$. Let us consider
a slightly perturbed solution $\phi=\phi_{0}+\pi$ of (\ref{EOM with Source})
with the source $J=J_{0}+\delta J$ and the original unperturbed metric
$g_{\mu\nu}\left(x\right)$. The equation of motion for $\pi$ is
then \begin{equation}
\tilde{G}^{\mu\nu}\nabla_{\mu}\nabla_{\nu}\pi+\varepsilon_{,\phi\phi}\pi+\varepsilon_{,\phi X}\delta X+\delta\tilde{G}^{\mu\nu}\nabla_{\mu}\nabla_{\nu}\phi_{0}=\delta J,\label{eom for pi}\end{equation}
where\[
\delta X=\nabla_{\nu}\phi_{0}\nabla^{\nu}\pi\text{ \thinspace\thinspace\,\, and}\text{ \thinspace\thinspace}\delta\tilde{G}^{\mu\nu}=\frac{\partial\tilde{G}^{\mu\nu}}{\partial\phi}\pi+\frac{\partial\tilde{G}^{\mu\nu}}{\partial\nabla_{\alpha}\phi}\nabla_{\alpha}\pi.\]
This equation can be written as \begin{equation}
\tilde{G}^{\mu\nu}\nabla_{\mu}\nabla_{\nu}\pi+V^{\mu}\nabla_{\mu}\pi+\tilde{M}^{2}\pi=\delta J,\label{EOM perturbations}\end{equation}
where \begin{equation}
V^{\mu}\left(x\right)\equiv\frac{\partial\tilde{G}^{\alpha\beta}}{\partial\nabla_{\mu}\phi}\nabla_{\alpha}\nabla_{\beta}\phi_{0}+\varepsilon_{,\phi X}\nabla^{\mu}\phi_{0},\label{vector}\end{equation}
and \begin{equation}
\tilde{M}^{2}\left(x\right)\equiv\frac{\partial\tilde{G}^{\alpha\beta}}{\partial\phi}\nabla_{\alpha}\nabla_{\beta}\phi_{0}+\varepsilon_{,\phi\phi}.\label{mtilda}\end{equation}
Considering the eikonal (or short wavelength) approximation \cite{LL}
we have $\pi\left(x\right)=A\left(x\right)\exp i\omega S\left(x\right)$,
where $\omega$ is a large dimensionless parameter and the amplitude
$A\left(x\right)$ is a slowly varying function. In the limit $\omega\rightarrow\infty$
the terms containing no second derivatives, $V^{\mu}\left(x\right)\nabla_{\mu}\pi$
and $\tilde{M}^{2}\left(x\right)\pi$, become unimportant and~(\ref{EOM perturbations})
becomes \begin{equation}
\tilde{G}^{\mu\nu}\partial_{\mu}S\partial_{\nu}S=0.\label{Eikonal}\end{equation}
The equation of motion in the eikonal approximation (\ref{Eikonal})
is conformally invariant. The surfaces of constant eikonal $S$ (constant
phase) correspond to the wave front (characteristic surface) in spacetime.
Thus the 1-form $\partial_{\mu}S$ is orthogonal to the characteristic
surface. The influence cone at point $P$ is formed by the propagation
vectors $N^{\mu}$ tangential to the characteristic surface $N^{\mu}\partial_{\mu}S=0$
and positive projection on the time direction. Using~(\ref{Eikonal})
one can chose $N^{\mu}=\tilde{G}^{\mu\nu}\partial_{\nu}S$ and verify
that this vectors are tangential to the characteristic surface. The
metric $\tilde{G}^{\mu\nu}$ has an inverse $\tilde{G}_{\mu\nu}^{-1}$
due to the requirement of hyperbolicity (Lorentzian signature of $\tilde{G}^{\mu\nu}$).
Therefore $\partial_{\nu}S=\tilde{G}_{\mu\nu}^{-1}N^{\mu}$ and we
obtain the equation for the influence cone in the form \[
\tilde{G}_{\mu\nu}^{-1}N^{\mu}N^{\nu}=0.\]
Thus the metric $\tilde{G}_{\mu\nu}^{-1}$ governs the division of
acoustic spacetime into past, future and inaccessible \textquotedblleft
spacelike\textquotedblright\ regions (or in other words this metric
yields the notion of causality). It is well known that this division
is invariant under conformal transformations. From action (\ref{Spi})
for perturbations $\pi$, which we derive in Appendix \ref{AppB}, it follows
that in four dimensions it is natural to consider a conformally transformed
metric $G_{\mu\nu}^{-1}=\left(\mathcal{L}_{,X}^{2}/c_{s}\right)\tilde{G}_{\mu\nu}^{-1}$.
Using this metric from (\ref{Gdown}) one obtains\[
G_{\mu\nu}^{-1}N^{\mu}N^{\nu}=\frac{\mathcal{L}_{,X}}{c_{s}}g_{\mu\nu}N^{\mu}N^{\nu}-c_{s}\mathcal{L}_{,XX}\left(\nabla_{\mu}\phi N^{\mu}\right)^{2}.\]
Therefore \[
g_{\mu\nu}N^{\mu}N^{\nu}=c_{s}^{2}\left(\frac{\mathcal{L}_{,XX}}{\mathcal{L}_{,X}}\right)\left(\nabla_{\mu}\phi N^{\mu}\right)^{2},\]
and if $\mathcal{L}_{,XX}/\mathcal{L}_{,X}$ is negative, then $g_{\mu\nu}N^{\mu}N^{\nu}<0$,
that is, $N^{\mu}$ is spacelike and the cone of influence on this
background is larger than the light cone: the wave front (or signal)
velocity is larger then the speed of light. Note that this is a coordinate
independent statement.
\section{Action for perturbations}
\label{AppB}
Here we sketch the derivation of action (\ref{Spi}) for $\pi$ in
the spacetime of arbitrary dimension $N>2$. First of all we would
like to investigate whether there exists a metric $G^{\mu\nu}$ for
which the equation of motion for perturbations $\pi$ takes a canonical
(Klein-Gordon) form \begin{equation}
G^{\mu\nu}D_{\mu}D_{\nu}\pi+M_{\text{eff}}^{2}\pi=\delta I,\label{Canonical}\end{equation}
 where $D_{\mu}$ is a covariant derivative with associated with the
new metric $G^{\mu\nu}$: $D_{\mu}G^{\alpha\beta}=0$. Note that the
equations of motion (\ref{eom for pi}) and (\ref{Canonical}) should
have the same influence cone structure. Thus the metrics $G^{\mu\nu}$
and $\tilde{G}^{\mu\nu}$ must be related by conformal transformation
and if it is really possible to rewrite (\ref{eom for pi}) in canonical
form, then there must exist $\Omega\left(\phi_{0},X_{0}\right),$
such that \begin{equation}
G^{\mu\nu}=\Omega\tilde{G}^{\mu\nu}.\label{def Omega}\end{equation}
 Therefore our first task is to find $\Omega\left(\phi_{0},X_{0}\right)$.
Note that this method makes sense for the dimensions $D>2$ only.
That happens because in $D=2$ all metrics are conformally equivalent
to $\eta_{\mu\nu}$ and the wave equation is conformally invariant,
see e.g. Ref.~\cite{Wald}, P. 447. Let us define the following covariant
derivative\begin{equation}
D_{\mu}A_{\nu}=\nabla_{\mu}A_{\nu}-L_{\mu\nu}^{\lambda}A_{\lambda}\label{DA}\end{equation}
 which is compatible with the new metric whereas $\nabla_{\mu}A_{\nu}=\partial_{\mu}A_{\nu}-\Gamma_{\mu\nu}^{\lambda}A_{\lambda}$
denotes the standard covariant derivative associated with the gravitational
metric: $\nabla_{\mu}g^{\alpha\beta}=0$, as usual. Note, that the
tensor $L_{\mu\nu}^{\lambda}$ introduced in (\ref{DA}) is the difference
of the Christoffel symbols corresponding to the effective and gravitational
metrics. Comparing (\ref{eom for pi}) and (\ref{Canonical}) we infer
that\[
\Omega\tilde{G}^{\mu\nu}D_{\mu}D_{\nu}\pi+M_{\text{eff}}^{2}\pi=\Omega\tilde{G}^{\mu\nu}\nabla_{\mu}\nabla_{\nu}\pi-\Omega\tilde{G}^{\mu\nu}L_{\mu\nu}^{\lambda}\nabla_{\lambda}\pi+M_{\text{eff}}^{2}\pi\]
 must be equal (up to a multiplication by a scalar function $\Omega$)
to the l.h.s of (\ref{EOM perturbations}). These can be true only
if the following condition holds\begin{equation}
\tilde{G}^{\mu\nu}L_{\mu\nu}^{\lambda}=-V^{\lambda},\label{eq:condition for L}\end{equation}
 where $V^{\lambda}$ is defined in (\ref{vector}). When this condition
is satisfied we can always make the redefinition \[
M_{\text{eff}}^{2}=\Omega\tilde{M}^{2}\text{ \,\, and \,\,}\delta I=\Omega\delta J,\]
 where $\tilde{M}^{2}$ is defined in (\ref{mtilda}). The connection
$L_{\mu\nu}^{\lambda}$ depends on the unknown function $\Omega$
(and its derivatives) which has to be obtained form (\ref{eq:condition for L}).
To solve (\ref{eq:condition for L}) it is convenient to multiply
its both sides by $\Omega$. Then using (\ref{vector}) and (\ref{def Omega})
this condition takes the form: \begin{equation}
G^{\mu\nu}L_{\mu\nu}^{\lambda}=-\Omega\left(\frac{\partial\tilde{G}^{\alpha\beta}}{\partial\nabla_{\lambda}\phi}\nabla_{\alpha}\nabla_{\beta}\phi_{0}+\varepsilon_{,\phi X}\nabla^{\lambda}\phi_{0}\right).\label{for Omega}\end{equation}
 Let us now solve (\ref{for Omega}) with respect to $\Omega$. In
complete analogy with the formula (86,6) from Ref.~\cite{LL} we
have \begin{equation}
G^{\mu\nu}L_{\mu\nu}^{\lambda}=-\frac{1}{\sqrt{-G}}\nabla_{\alpha}\left(\sqrt{-G}G^{\alpha\lambda}\right),\label{Cristofel trace}\end{equation}
 where $\sqrt{-G}=\sqrt{-\text{det}G_{\mu\nu}^{-1}}=\Omega^{-D/2}\sqrt{-\text{det}\tilde{G}_{\alpha\beta}^{-1}}$,
and $D$ is the number of dimensions of the spacetime. Using the formula
(B14) from Ref.~\cite{Halo} one obtains \begin{eqnarray}
\text{det}\tilde{G}^{\alpha\beta}=\left(\mathcal{L}_{,X}\right)^{D}c_{s}^{-2}\text{det}\left(g^{\mu\nu}\right), & \text{and} & \text{det}\tilde{G}_{\alpha\beta}^{-1}=\left(\mathcal{L}_{,X}\right)^{-D}c_{s}^{2}\text{det}\left(g_{\mu\nu}\right).\label{eq: Determinants}\end{eqnarray}
 Finally we arrive to the relation,\begin{equation}
\sqrt{-G}=c_{s}\sqrt{-g}\left(\Omega\mathcal{L}_{,X}\right)^{-D/2}.\end{equation}
 It is convenient to introduce the auxiliary function \begin{equation}
F=c_{s}\left(\Omega\mathcal{L}_{,X}\right)^{-D/2}\Omega.\end{equation}
 and then using~(\ref{Cristofel trace}), we can rewrite equation
(\ref{for Omega}) as: \begin{equation}
\nabla_{\alpha}\left(F\tilde{G}^{\alpha\lambda}\right)=F\left(\frac{\partial\tilde{G}^{\alpha\beta}}{\partial\nabla_{\lambda}\phi}\nabla_{\alpha}\nabla_{\beta}\phi_{0}+\varepsilon_{,\phi X}\nabla^{\lambda}\phi_{0}\right).\end{equation}
 Differentiating the metric $\tilde{G}^{\alpha\lambda}$ from the
l.h.s. of the last equation in accordance with the chain rule we find:
\begin{equation}
\tilde{G}^{\alpha\lambda}\nabla_{\alpha}F=F\left(\left(\frac{\partial\tilde{G}^{\alpha\beta}}{\partial\nabla_{\lambda}\phi}-\frac{\partial\tilde{G}^{\alpha\lambda}}{\partial\nabla_{\beta}\phi}\right)\nabla_{\alpha}\nabla_{\beta}\phi_{0}-\left(\frac{\partial\tilde{G}^{\alpha\lambda}}{\partial\phi}-\varepsilon_{,\phi X}g^{\lambda\alpha}\right)\nabla_{\alpha}\phi_{0}\right).\label{for F}\end{equation}
 Further we obtain\begin{equation}
\frac{\partial\tilde{G}^{\alpha\lambda}}{\partial\phi}\nabla_{\alpha}\phi_{0}=\left(\mathcal{L}_{,X\phi}+2X\mathcal{L}_{,XX\phi}\right)\nabla^{\lambda}\phi_{0}=\varepsilon_{,\phi X}\nabla^{\lambda}\phi_{0}.\end{equation}
 For the first term in the brackets in~(\ref{for F}) we have:\begin{equation}
\frac{\partial\tilde{G}^{\alpha\beta}}{\partial\nabla_{\lambda}\phi}=\mathcal{L}_{,XX}\left(g^{\alpha\beta}\nabla^{\lambda}\phi_{0}+g^{\lambda\alpha}\nabla^{\beta}\phi_{0}+g^{\lambda\beta}\nabla^{\alpha}\phi_{0}\right)+\mathcal{L}_{,XXX}\nabla^{\alpha}\phi_{0}\nabla^{\beta}\phi_{0}\nabla^{\lambda}\phi_{0},\end{equation}
 and therefore\begin{equation}
\frac{\partial\tilde{G}^{\alpha\beta}}{\partial\nabla_{\lambda}\phi}-\frac{\partial\tilde{G}^{\alpha\lambda}}{\partial\nabla_{\beta}\phi}=0.\end{equation}
 Thus the r.h.s. of (\ref{for F}) identically vanishes. Note that
there exists the inverse matrix $\tilde{G}_{\alpha\lambda}^{-1}$
to $\tilde{G}^{\alpha\lambda}$. Therefore from (\ref{for F}) we
conclude that $\nabla_{\alpha}F=0$ or $F=const$ on all backgrounds
and for all theories. Considering the linear case, $\mathcal{L}\left(\phi,X\right)=X-V(\phi),$
we infer that $F=c_{s}\left(\Omega\mathcal{L}_{,X}\right)^{-D/2}\Omega=1$
or \begin{equation}
\Omega=\left(c_{s}\mathcal{L}_{,X}^{-D/2}\right)^{1/\left(D/2-1\right)}.\end{equation}
 Having calculated $\Omega$ we can formulate the main result of this
Appendix as follows: the action from which one can obtain the equation
of motion in the canonical Klein-Gordon form (\ref{Canonical}) is\begin{equation}
S_{\pi}=\frac{1}{2}\int d^{D}x\sqrt{-G}\,\left[G^{\mu\nu}\partial_{\mu}\pi\partial_{\nu}\pi-M_{\text{eff}}^{2}\pi^{2}+2\pi\delta I\right],\label{Spi N}\end{equation}
 where the emergent metric $G^{\mu\nu}$ is the conformally transformed
eikonal metric $\tilde{G}^{\mu\nu},$ defined in~(\ref{Metric}),
\begin{equation}
G^{\mu\nu}\equiv\left(c_{s}\mathcal{L}_{,X}^{-D/2}\right)^{1/\left(D/2-1\right)}\tilde{G}^{\mu\nu}=\left(\frac{c_{s}}{\mathcal{L}_{,X}}\right)^{1/\left(D/2-1\right)}\left[g^{\mu\nu}+\left(\frac{\mathcal{L}_{,XX}}{\mathcal{L}_{,X}}\right)\nabla^{\mu}\phi\nabla^{\nu}\phi\right].\label{Gup N}\end{equation}
 The inverse metric $G_{\mu\nu}^{-1}$ can be easily calculated using
the ansatz $G_{\mu\nu}^{-1}=\alpha g_{\mu\nu}+\beta\nabla_{\mu}\phi_{0}\nabla_{\nu}\phi_{0}$
and is given by the formula\begin{equation}
G_{\mu\nu}^{-1}=\left(\frac{c_{s}}{\mathcal{L}_{,X}}\right)^{-1/\left(D/2-1\right)}\left[g^{\mu\nu}-c_{s}^{2}\left(\frac{\mathcal{L}_{,XX}}{\mathcal{L}_{,X}}\right)\nabla^{\mu}\phi_{0}\nabla^{\nu}\phi_{0}\right].\label{Gdown N}\end{equation}
 Finally the effective mass is \begin{equation}
M_{\text{eff}}^{2}=\left(c_{s}\mathcal{L}_{,X}^{-N/2}\right)^{1/\left(D/2-1\right)}\left[2X\mathcal{L}_{,X\phi\phi}-\mathcal{L}_{,\phi\phi}+\frac{\partial\tilde{G}^{\mu\nu}}{\partial\phi}\nabla_{\mu}\nabla_{\nu}\phi_{0}\right],\label{Mass N}\end{equation}
 and the effective source for perturbations is given by \begin{equation}
\delta I=\left(c_{s}\mathcal{L}_{,X}^{-D/2}\right)^{1/\left(D/2-1\right)}\delta J.\label{eq: Source rescalation}\end{equation}
 For the reference we also list the formula \begin{equation}
\sqrt{-G}=\sqrt{-g}\left(\frac{\mathcal{L}_{,X}^{D}}{c_{s}^{2}}\right)^{1/\left(D-2\right)}.\label{root of G}\end{equation}
\section{Action for the cosmological perturbations}
\label{AppC}
Here we compare the action (\ref{Spi}) with the action for scalar
cosmological perturbations from Refs.~\cite{Mukhanov BOOK,GarMukh}.
In particular we show that cosmological perturbations propagate in
the metric (\ref{Gdown}) but have an effective mass different from
(\ref{Mass}). Finally we derive the generally covariant action for
the scalar cosmological perturbations.

To begin with let us consider the action (\ref{Spi}) for a perturbations
$\pi\left(\eta,\mathbf{x}\right)$ around a homogeneous background
$\phi(\eta)$ in the spatially flat Friedmann universe\begin{equation}
ds^{2}=g_{\mu\nu}dx^{\mu}dx^{\nu}=a^{2}\left(\eta\right)\left(d\eta^{2}-d\mathbf{x}^{2}\right)=a^{2}\left(\eta\right)\eta_{\mu\nu}dx^{\mu}dx^{\nu}\label{conformal Friedmann}\end{equation}
 where $\eta$ is the conformal time $\eta=\int dt/a\left(t\right)$
and $\eta_{\mu\nu}$ is the standard Minkowski metric. Using Eq.~(\ref{cs})
and Eq.~(\ref{conformal Friedmann}) one can calculate the effective
line element (\ref{interval}): \begin{eqnarray}
dS^{2} & = & G_{\mu\nu}^{-1}dx^{\mu}dx^{\nu}=\frac{\mathcal{L}_{,X}}{c_{s}}\left[ds^{2}-a^{2}c_{s}^{2}\left(\frac{\mathcal{L}_{,XX}}{\mathcal{L}_{,X}}\right)2Xd\eta^{2}\right]=\notag\\
 &  & =\frac{\mathcal{L}_{,X}}{c_{s}}a^{2}\left(c_{s}^{2}d\eta^{2}-d\mathbf{x}^{2}\right)\equiv c_{s}A^{2}\left(c_{s}^{2}d\eta^{2}-d\mathbf{x}^{2}\right).\label{eq: Conformal effective line el}\end{eqnarray}
 where we have introduced the convenient variable \begin{equation}
A\equiv\sqrt{\varepsilon_{,X}}a.\label{A}\end{equation}
 Note that for the models respecting the NEC ($\mathcal{L}_{,X}\geq0$)
the hyperbolicity condition (\ref{HYPER}) requires $\varepsilon_{,X}>0$
and therefore $A$ is always well defined. The factor $\sqrt{-G}$
can be then calculated either from the last expression above (\ref{eq: Conformal effective line el})
or from the general expression (\ref{root of G}): \begin{equation}
\sqrt{-G}=\frac{\mathcal{L}_{,X}^{2}}{c_{s}}a^{4}=c_{s}^{3}A^{4}.\end{equation}
 Using the formulas (\ref{Gup}) and (\ref{cs}) we calculate the
kinetic term\begin{equation}
G^{\mu\nu}\partial_{\mu}\pi\partial_{\nu}\pi=\left(c_{s}a^{2}\mathcal{L}_{,X}\right)^{-1}\left(\left(\pi^{\prime}\right)^{2}-c_{s}^{2}(\vec{\nabla}\pi)^{2}\right).\end{equation}
 Thus in the case when the perturbations $\pi$ do not influence the
metric $g_{\mu\nu}$ the action (\ref{Spi}) takes the form\begin{equation}
S_{\pi}=\frac{1}{2}\int d^{3}xd\eta\,\left[a^{2}\varepsilon_{,X}\left(\left(\pi^{\prime}\right)^{2}-c_{s}^{2}(\vec{\nabla}\pi)^{2}\right)-M_{\text{eff}}^{2}\frac{\mathcal{L}_{,X}^{2}}{c_{s}}a^{4}\pi^{2}\right],\end{equation}
 here we have used the definitions of the sound speed (\ref{cs})
and energy density (\ref{e}). It is convenient to introduce the canonical
normalization for the perturbations. This is achieved by the following
field redefinition: \begin{equation}
\nu=\sqrt{\varepsilon_{,X}}a\pi=\pi A.\end{equation}
 Finally integrating by parts and dropping the total derivative terms
we obtain the following {}``canonical'' action \begin{equation}
S_{\pi}=\frac{1}{2}\int d^{3}xd\eta\,\left[(\nu^{\prime})^{2}-c_{s}^{2}(\vec{\nabla}\nu)^{2}-m_{\text{eff}}^{2}\nu^{2}\right]\label{Action cosmology our}\end{equation}
 where the new effective mass $m_{\text{eff}}$ is given by the following
expression\begin{equation}
m_{\text{eff}}^{2}=M_{\text{eff}}^{2}\frac{\sqrt{-G}}{A^{2}}-\frac{A^{\prime\prime}}{A}=\frac{a^{2}}{\varepsilon_{,X}}\left[\varepsilon_{,\phi\phi}+\frac{\partial\tilde{G}^{\mu\nu}}{\partial\phi}\nabla_{\mu}\nabla_{\nu}\phi_{0}\right]-\frac{\left(\sqrt{\varepsilon_{,X}}a\right)^{\prime\prime}}{\sqrt{\varepsilon_{,X}}a}.\label{eff mass for pi}\end{equation}
 or in other terms\begin{equation}
m_{\text{eff}}^{2}=\frac{1}{\varepsilon_{,X}}\left[\varepsilon_{,X\phi}\phi^{\prime\prime}+\mathcal{H}\phi^{\prime}\left(3p_{,X\phi}-\varepsilon_{,X\phi}\right)+\varepsilon_{,\phi\phi}a^{2}\right]-\frac{\left(\sqrt{\varepsilon_{,X}}a\right)^{\prime\prime}}{\sqrt{\varepsilon_{,X}}a}.\end{equation}
 Now let us consider the case of cosmological perturbations in the
case where the field $\phi$ is responsible for the dynamics of the
Friedmann universe. Following \cite{GarMukh,Mukhanov BOOK} one introduces
a canonical variable $\upsilon$\begin{equation}
\upsilon\equiv\sqrt{\varepsilon_{,X}}a\left(\delta\phi+\frac{\phi^{\prime}}{\mathcal{H}}\Psi\right)=A\left(\delta\phi+\frac{\phi^{\prime}}{\mathcal{H}}\Psi\right),\end{equation}
 and a convenient auxiliary variable $z$\begin{equation}
z\equiv\frac{\phi^{\prime}}{\mathcal{H}}\sqrt{\varepsilon_{,X}}a=\frac{\phi^{\prime}}{\mathcal{H}}A,\end{equation}
 where $\delta\phi$ is the gauge invariant perturbation of the scalar
field, $\mathcal{H}\equiv a^{\prime}/a$ and $\Psi=\Phi$ is the gauge
invariant Newtonian potential. Using this notation the action for
scalar cosmological perturbations takes the form:\begin{equation}
S_{\text{cosm}}=\frac{1}{2}\int d^{3}xd\eta\,\left[(\upsilon^{\prime})^{2}-c_{s}^{2}(\vec{\nabla}\upsilon)^{2}-m_{\text{cosm}}^{2}\upsilon^{2}\right]\label{Action GarrMukh}\end{equation}
 where \begin{equation}
m_{\text{cosm}}^{2}\equiv-\frac{z^{\prime\prime}}{z}.\end{equation}
 It is easily to check that for all cases besides canonical field
without potential $\mathcal{L}(\phi,X)\equiv X$\begin{equation}
m_{\text{cosm}}^{2}\neq m_{\text{eff}}^{2}.\end{equation}
 However, comparing the action (\ref{Action cosmology our}) with
\ref{Action GarrMukh} one arrives to conclusion that the cosmological
perturbations propagate in the same metric (\ref{Gup}), (\ref{Gdown}).
Further one can introduce the notation $\overline{\delta\phi}$ for
the sometimes so-called {}``scalar perturbations on the spatially
flat slicing'' \begin{equation}
\overline{\delta\phi}\equiv\delta\phi+\frac{\phi^{\prime}}{\mathcal{H}}\Psi.\end{equation}
 For this scalar field the action for cosmological perturbations (\ref{Action GarrMukh})
takes the form \begin{equation}
S_{\text{cosm}}=\frac{1}{2}\int d^{4}x\sqrt{-G}\,\left[G^{\mu\nu}\partial_{\mu}\overline{\delta\phi}\partial_{\nu}\overline{\delta\phi}-M_{\text{cosm}}^{2}\overline{\delta\phi}^{2}\right],\label{eq: Scosmo}\end{equation}
 thus the cosmological perturbations $\overline{\delta\phi}$ live
in the emergent acoustic spacetime with the metric (\ref{Gup}), (\ref{Gdown}).
Similarly as we have calculated in (\ref{eff mass for pi}) we have\begin{equation}
M_{\text{cosm}}^{2}\frac{\sqrt{-G}}{A^{2}}-\frac{A^{\prime\prime}}{A}=M_{\text{cosm}}^{2}a^{2}\mathcal{L}_{,X}c_{s}-\frac{\left(\sqrt{\varepsilon_{,X}}a\right)^{\prime\prime}}{\sqrt{\varepsilon_{,X}}a}=-\frac{z^{\prime\prime}}{z}\end{equation}
 after some algebra the last expression reduces to \begin{equation}
\chi^{\prime\prime}+2\left(\frac{A^{\prime}}{A}\right)\chi^{\prime}+A^{2}\left(M_{\text{cosm}}^{2}c_{s}^{3}\right)\chi=0\label{Klein Gordon for y}\end{equation}
 where we have introduced a new auxiliary field \begin{equation}
\chi\left(\eta\right)\equiv\frac{\phi^{\prime}}{\mathcal{H}}=\left(\frac{3}{8\pi G_{N}}\right)^{1/2}\sqrt{\frac{2X}{\varepsilon}}.\end{equation}
 The equation (\ref{Klein Gordon for y}) is in turn the Klein-Gordon
equation \begin{equation}
\left(\Box_{\overline{g}}+\left(M_{\text{cosm}}^{2}c_{s}^{3}\right)\right)\chi=0\end{equation}
 for the field $\chi$ in the metric $\overline{g}_{\mu\nu}\equiv A^{2}\eta_{\mu\nu}=\varepsilon_{,X}g_{\mu\nu}$
conformally related to the gravitational metric $g_{\mu\nu}$. Thus
we have \begin{equation}
M_{\text{cosm}}^{2}=-c_{s}^{-3}\chi^{-1}\Box_{\overline{g}}\,\chi.\label{cosmo mass for gconform}\end{equation}
 One can rewrite this formula in terms of the gravitational metric
$g_{\mu\nu}$. Using the rules of the conformal transformations we
have \begin{eqnarray}
\Box_{\overline{g}}\chi & = & \frac{1}{\sqrt{-\overline{g}}}\nabla_{\mu}\left(\sqrt{-\overline{g}}\overline{g}^{\mu\nu}\nabla_{\nu}\chi\right)=\frac{1}{\varepsilon_{,X}^{2}}\frac{1}{\sqrt{-g}}\nabla_{\mu}\left(\varepsilon_{,X}\sqrt{-g}g^{\mu\nu}\nabla_{\nu}\chi\right)=\\
 &  & =-\nabla^{\mu}\chi\nabla_{\mu}\varepsilon_{,X}^{-1}+\varepsilon_{,X}^{-1}\Box_{g}\chi\end{eqnarray}
 Thus the effective mass for cosmological perturbations $\overline{\delta\phi}$
is \begin{equation}
M_{\text{cosm}}^{2}=-c_{s}^{-3}\varepsilon_{,X}^{-1}\left(\sqrt{\frac{\varepsilon}{X}}\Box_{g}\sqrt{\frac{X}{\varepsilon}}+\nabla_{\mu}\ln\left(\varepsilon_{,X}\right)\nabla^{\mu}\ln\sqrt{\frac{X}{\varepsilon}}\right).\label{cosmo mass for normal g}\end{equation}
 Note that in the case of canonical kinetic terms $\mathcal{L}(\phi,X)=X-V\left(\phi\right)$
the last expression for $M_{\text{cosm}}^{2}$ the simplifies to \begin{equation}
M_{\text{cosm},\text{canonical}}^{2}=-\left(w+1\right)^{-1/2}\Box_{g}\left(w+1\right)^{1/2}.\end{equation}
 where $w=p/\varepsilon$ is the equation of state parameter. In particular
for the universe filled with the massless canonical scalar field $M_{\text{cosm},}=0$.
\section{Effective Hydrodynamics}
\label{AppD}
It is well-known that for timelike $\nabla_{\nu}\phi$ ($X>0$ in
our signature) one can employ the hydrodynamic approach to describe
the system with the action (\ref{action}). To do this one need to
introduce a four-velocity as follows: \begin{equation}
u_{\mu}\equiv\frac{\nabla_{\mu}\phi}{\sqrt{2X}}.\label{u}\end{equation}
 Using (\ref{u}) the energy momentum tensor (\ref{EMT}) tensor can
be rewritten in the perfect fluid form:\[
T_{\mu\nu}=\left(\varepsilon+p\right)u_{\mu}u_{\nu}-pg_{\mu\nu},\]
 where the pressure coincides with the Lagrangian density, $p=\mathcal{L}(X,\phi),$
and the energy density is\begin{equation}
\varepsilon\left(X,\phi\right)=2Xp_{,X}-p.\label{e}\end{equation}
 The sound speed (\ref{cs}) can be expressed \cite{GarMukh} as \begin{equation}
c_{s}^{2}=\frac{p_{,X}}{\varepsilon_{,X}}=\left(\frac{\partial p}{\partial\varepsilon}\right)_{\phi}.\label{sound in kessence}\end{equation}
 In what follows we restrict ourselves to the class of Lagrangians
which do not depend of $\phi$ explicitly, $p=p\left(X\right)$ and
in addition we require that $X>0$. This class of models is equivalent
to perfect fluid models with zero vorticity and with the pressure
being a function of the energy density only, $p=p(\epsilon)$. Then
the expressions (\ref{cs}) or (\ref{sound in kessence}) coincide
with the usual definition of the sound speed for the perfect fluid:
$c_{s}^{2}=\partial p/\partial\varepsilon.$ Apart from the energy
density $\varepsilon$ and pressure $p$ one can also formally introduce
the {}``concentration of particles'': \[
n\equiv\exp\left(\int\frac{d\varepsilon}{\varepsilon+p(\varepsilon)}\right)=\sqrt{X}p_{,X}.\]
 and the enthalpy\[
h\equiv\frac{\varepsilon+p}{n}=2\sqrt{X}.\]
 In particular the equation of motion (\ref{EOM}) takes the form
of the particle number conservation law: $\nabla_{\mu}\left(nu^{\mu}\right)=0$.
Using these definitions we can rewrite the induced metric metric $G^{\mu\nu}$
and its inverse in terms of hydrodynamic quantities only:\begin{eqnarray}
G^{\mu\nu} & = & \frac{hc_{s}}{2n}\left[g^{\mu\nu}-\left(1-c_{s}^{-2}\right)u^{\mu}u^{\nu}\right],\label{G}\\
G_{\mu\nu}^{-1} & = & \frac{2n}{hc_{s}}\left[g_{\mu\nu}-\left(1-c_{s}^{2}\right)u_{\mu}u_{\nu}\right].\end{eqnarray}
 To our best knowledge these metrics (\ref{G}) along with an action
for the velocity potentials were introduced for the first time in
\cite{Moncrief}, where the accretion of the perfect fluid onto black
hole was studied. As it follows from the derivation in Appendix \ref{AppB},
the metric (\ref{Gup N}) and the action (\ref{Spi N}) derived in
our paper are applicable in the more general case of arbitrary nonlinear
scalar field theories $\mathcal{L}\left(X,\phi\right)$ and for all
possible (not only timelike $X_{0}>0$) backgrounds produced by any
external sources. Note that the scalar field theory with Lagrangian
$\mathcal{L}\left(X,\phi\right),$ which explicitly depends on $\phi,$
is not equivalent to the isentropic hydrodynamics, because $\phi$
and $X$ are independent and therefore the pressure cannot be expressed
though $\varepsilon$ only.
\section{Green functions for a moving spacecraft}
\label{AppE}
Here we calculate the retarded Green's function for a moving spacecraft
in the case of three spatial dimensions. First we calculate the retarded
Green's function in the preferred (rest) frame and then we perform
the Lorentz boost (with the invariant speed $c$) for the solution.
We compare the result with one obtained by the direct calculation
of Green's function for the Eq.~(\ref{EOM normal Lorentz}). We will
need the following formulas (Gradshtein, Ryzhik, p.750):\begin{eqnarray}
\int_{a}^{\infty}J_{0}\left(b\sqrt{x^{2}-a^{2}}\right)\sin\left(cx\right) & = & \frac{\cos\left(a\sqrt{c^{2}-b^{2}}\right)}{\sqrt{c^{2}-b^{2}}},\quad\mathrm{for}\quad0<b<c\label{gr1}\\
 & = & 0,\quad\mathrm{for}\quad0<c<b\label{gr2}\\
\int_{a}^{\infty}J_{0}\left(b\sqrt{x^{2}-a^{2}}\right)\cos\left(cx\right) & = & -\frac{\sin\left(a\sqrt{c^{2}-b^{2}}\right)}{\sqrt{c^{2}-b^{2}}},\quad\mathrm{for}\quad0<b<c\label{gr3}\\
 & = & \frac{\exp\left(-a\sqrt{b^{2}-c^{2}}\right)}{\sqrt{b^{2}-c^{2}}},\quad\mathrm{for}\quad0<c<b\label{gr4}\\
\int_{0}^{a}J_{0}\left(b\sqrt{a^{2}-x^{2}}\right)\cos\left(cx\right) & = & \frac{\sin\left(a\sqrt{c^{2}+b^{2}}\right)}{\sqrt{c^{2}+b^{2}}},\quad\mathrm{for}\quad0<b\label{gr5}\end{eqnarray}
 In the preferred frame the Green function is (see e.g. \cite{Vladimirov})
\begin{equation}
G_{R}^{\text{rf}}(t,x)=\frac{\theta\left(t\right)}{2c_{s}\pi}\delta\left(c_{s}^{2}t^{2}-|x|^{2}\right).\end{equation}
 Performing the Lorentz transformation $x=\gamma\left(x^{\prime}+vt^{\prime}\right)$,
$t=\gamma\left(t^{\prime}+vx^{\prime}\right)$, where $\gamma=\left(1-v^{2}\right)^{-1/2}$
we find the Green function in the moving frame: \begin{equation}
G_{R}^{\text{rf}}(t^{\prime},x^{\prime})=\frac{\theta\left(t^{\prime}+vx^{\prime}\right)}{2c_{s}\pi}\delta\left[\gamma^{2}\left(c_{s}^{2}\left(t^{\prime}+vx^{\prime}\right)^{2}-\left(x^{\prime}+vt^{\prime}\right)^{2}\right)-y^{2}-z^{2}\right].\label{green1}\end{equation}
 We need to calculate the Fourier transform to the function (\ref{green1}).
It is convenient to shift $x^{\prime}$ as follows:\begin{equation}
x^{\prime}=\overline{x}-vt^{\prime}\left(\frac{1-c_{s}^{2}}{1-c_{s}^{2}v^{2}}\right).\end{equation}
 Then the argument of the delta-function in (\ref{green1}) can be
rewritten as\[
\gamma^{2}\left(c_{s}^{2}\left(t^{\prime}+vx^{\prime}\right)^{2}-\left(x^{\prime}+vt^{\prime}\right)^{2}\right)-y^{2}-z^{2}=\alpha c_{s}^{2}t^{\prime}{}^{2}-\alpha^{-1}\overline{x}^{2}-y^{2}-z^{2},\]
 where\begin{equation}
\alpha=\frac{1-v^{2}}{1-c_{s}^{2}v^{2}}.\end{equation}
 Now we are ready to proceed with the Fourier transform of (\ref{green1}):\begin{equation}
G_{R}^{\text{rf}}(t^{\prime},k^{\prime})=\frac{\text{e}^{i\varphi}}{2c_{s}\pi}\int_{-\infty}^{\infty}d\overline{x}dydz\,\theta\left(t^{\prime}+vx^{\prime}\right)\,\delta\left(\alpha c_{s}^{2}t^{\prime}{}^{2}-\alpha^{-1}\overline{x}^{2}-y^{2}-z^{2}\right)\text{e}^{ik_{x^{\prime}}\overline{x}+ik_{y}y+ik_{z}z}\end{equation}
 where we introduced the notation:\begin{equation}
\varphi=-k_{x^{\prime}}vt^{\prime}\left(\frac{1-c_{s}^{2}}{1-c_{s}^{2}v^{2}}\right).\end{equation}
 Step-function in the integral implies that the integration over $\overline{x}$
is made from $x_{*}$ to $+\infty$:
\begin{equation}
G_{R}^{\mathrm{rf}}(t^{\prime},k^{\prime})=\frac{\text{e}^{i\varphi}}{2c_{s}\pi}\int_{x_{*}}^{\infty}d\overline{x}\int_{-\infty}^{\infty}dy\int_{-\infty}^{\infty}dz\,\delta\left(\alpha c_{s}^{2}t^{\prime}{}^{2}-\alpha^{-1}\overline{x}^{2}-y^{2}-z^{2}\right)\text{e}^{ik_{x^{\prime}}\overline{x}+ik_{y}y+ik_{z}z},\end{equation}
\begin{equation}
x_{*}=vt^{\prime}\left(\frac{1-c_{s}^{2}}{1-c_{s}^{2}v^{2}}\right)-\frac{t^{\prime}}{v}=-\frac{t^{\prime}}{v}\left(\frac{1-v^{2}}{1-c_{s}^{2}v^{2}}\right)=-\frac{\alpha}{v}t^{\prime}.\end{equation}
 Introducing $r\equiv\sqrt{y^{2}+z^{2}}$, $\phi$ as the angle between
the vectors $\left\{ k_{y},k_{z}\right\} $ and $\left\{ y,z\right\} $
and $k_{\bot}\equiv\sqrt{k_{y}^{2}+k_{z}^{2}}$ we obtain:\begin{equation}
G_{R}^{\mathrm{rf}}(t^{\prime},k^{\prime})=\frac{\text{e}^{i\varphi}}{2c_{s}\pi}\int_{x_{*}}^{\infty}d\overline{x}\int_{0}^{\infty}drr\int_{0}^{2\pi}d\phi\delta\left(\alpha c_{s}^{2}t^{\prime}{}^{2}-\alpha^{-1}\overline{x}^{2}-r^{2}\right)\text{e}^{ik_{x^{\prime}}\overline{x}+ik_{\bot}r\cos\phi}.\end{equation}
 Integrating over $r$ first gives:\begin{equation}
G_{R}^{\mathrm{rf}}(t^{\prime},k^{\prime})=\frac{\text{e}^{i\varphi}}{4c_{s}\pi}\int_{x_{*}}^{+\infty}d\overline{x}\int_{0}^{2\pi}d\phi\exp\left(ik_{x^{\prime}}\overline{x}+ik_{\bot}\sqrt{\alpha c_{s}^{2}t^{\prime}{}^{2}-\alpha^{-1}\overline{x}^{2}}\cos\phi\right)\label{int1}\end{equation}
for
\begin{equation}
\alpha c_{s}^{2}t^{\prime}{}^{2}-\alpha^{-1}\overline{x}^{2}>0,\label{condition}\end{equation}
 otherwise it is zero. Integrating (\ref{int1}) over $\phi$ we find:
\begin{equation}
G_{R}^{\mathrm{rf}}(t^{\prime},k^{\prime})=\frac{\text{e}^{i\varphi}}{2c_{s}}\int_{x_{*}}^{+\infty}d\overline{x}J_{0}\left(k_{\bot}\sqrt{\alpha c_{s}^{2}t^{\prime}{}^{2}-\alpha^{-1}\overline{x}^{2}}\right)\exp\left(ik_{x^{\prime}}\overline{x}\right),\label{int}\end{equation}
 where $J_{0}(x)$ is the Bessel function of the zeroth order. Now
we need to integrate the expression (\ref{int}) taking into account
the condition (\ref{condition}).We consider two cases separately:
the case of slow spacecraft, $v^{2}c_{s}^{2}<1$ ($\alpha>0$), and
the case of rapid spacecraft, $v^{2}c_{s}^{2}>1$ ($\alpha<0$).

For the slow spacecraft we easily obtain from (\ref{int}) and (\ref{condition}):
\begin{eqnarray*}
G_{R}^{\mathrm{rf}}(t^{\prime},k^{\prime}) & = & \frac{\text{e}^{i\varphi}}{2c_{s}}\theta(t^{\prime})\int_{-\alpha c_{s}t^{\prime}}^{\alpha c_{s}t^{\prime}}d\overline{x}\, J_{0}\left(k_{\bot}\sqrt{\alpha c_{s}^{2}t^{\prime}{}^{2}-\alpha^{-1}\overline{x}^{2}}\right)e^{ik_{x^{\prime}}\overline{x}}\\
 & = & \frac{\text{e}^{i\varphi}}{c_{s}}\theta(t^{\prime})\int_{0}^{\alpha c_{s}t^{\prime}}d\overline{x}\, J_{0}\left(\frac{k_{\bot}}{\sqrt{\alpha}}\sqrt{\alpha^{2}c_{s}^{2}t^{\prime}{}^{2}-\overline{x}^{2}}\right)\cos\left(k_{x^{\prime}}\overline{x}\right).\end{eqnarray*}
 Using (\ref{gr5}) we then find the Green's function for slow moving
spacecraft:\begin{eqnarray}
G_{R}^{\mathrm{rf}}(t^{\prime},k^{\prime}) & = & -\theta\left(t^{\prime}\right)\frac{i\text{e}^{i\varphi}}{2c_{s}\sqrt{k_{x^{\prime}}^{2}+k_{\bot}^{2}/\alpha}}\left(\text{e}^{i\alpha c_{s}t^{\prime}\sqrt{k_{x^{\prime}}^{2}+k_{\bot}^{2}/\alpha}}-\text{e}^{-i\alpha c_{s}t^{\prime}\sqrt{k_{x^{\prime}}^{2}+k_{\bot}^{2}/\alpha}}\right)\notag\\
 & = & \theta\left(t^{\prime}\right)\frac{1}{2ic_{s}}\left(k_{x^{\prime}}^{2}+k_{\bot}^{2}\frac{1-c_{s}^{2}v^{2}}{1-v^{2}}\right)^{-1/2}\left(\text{e}^{i\omega_{+}t^{\prime}}-\text{e}^{i\omega_{-}t^{\prime}}\right).\label{Retarded Green in slow rocket app}\end{eqnarray}

In the case of rapid spacecraft, $v^{2}c_{s}^{2}>1$ $\left(\alpha<0\right)$,
one can verify that $\alpha^{2}c_{s}^{2}t^{\prime}{}^{2}>x_{*}^{2}$
for any $t^{\prime}$. Thus (\ref{int}) along with (\ref{condition})
can be rewritten as:\begin{equation}
G_{R}^{\mathrm{rf}}(t^{\prime},k^{\prime})=\frac{\text{e}^{i\varphi}}{2c_{s}}\int_{\left|\alpha c_{s}t^{\prime}\right|}^{+\infty}d\overline{x}\, J_{0}\left(\frac{k_{\bot}}{\sqrt{\left|\alpha\right|}}\sqrt{\alpha^{2}c_{s}^{2}t^{\prime}{}^{2}-\overline{x}^{2}}\right)\left(\cos\left(k_{x^{\prime}}\overline{x}\right)+i\sin\left(k_{x^{\prime}}\overline{x}\right)\right).\end{equation}
 Using (\ref{gr2}) and (\ref{gr4}) for $k_{\bot}^{2}>\left|\alpha\right|k_{x^{\prime}}^{2}$
and (\ref{gr1}) and (\ref{gr3}) for $k_{\bot}^{2}<\left|\alpha\right|k_{x^{\prime}}^{2}$
we obtain in both cases:
\begin{equation}
G_{R}^{\mathrm{rf}}(t^{\prime},k^{\prime})=\frac{\text{e}^{i\varphi}}{2c_{s}}\frac{\exp\left(-\left|\alpha c_{s}t^{\prime}\right|\sqrt{k_{\bot}^{2}\left|\alpha\right|^{-1}-k_{x^{\prime}}^{2}}\right)}{\sqrt{k_{\bot}^{2}\left|\alpha\right|^{-1}-k_{x^{\prime}}^{2}}}.\end{equation}
 The last expression can be written as \begin{equation}
G_{R}^{\mathrm{rf}}(t^{\prime},k^{\prime})=-\frac{1}{2ic_{s}}\left(k_{x^{\prime}}^{2}+k_{\bot}^{2}\frac{1-c_{s}^{2}v^{2}}{1-v^{2}}\right)^{-1/2}\left(\theta\left(t^{\prime}\right)\text{e}^{i\omega_{+}t^{\prime}}+\theta\left(-t^{\prime}\right)\text{e}^{i\omega_{-}t^{\prime}}\right).\label{Fast moving green correct app}\end{equation}
 Thus the modes propagating in with \begin{equation}
k_{\bot}^{2}>k_{x^{\prime}}^{2}\left|\alpha\right|=k_{x^{\prime}}^{2}\left(\frac{1-v^{2}}{c_{s}^{2}v^{2}-1}\right)\end{equation}
 are exponentially suppressed. The singular directions $k_{\bot}^{2}=k_{x^{\prime}}^{2}\left|\alpha\right|$
are unphysical because they have measure zero in the integral. This
directions correspond to the sufficient but integrable singularities
in the Green function.

If the Green's function is calculated directly from the Eq.~(\ref{EOM normal Lorentz})
by means of standard approach then one can find, that the solution
is:\begin{eqnarray}
G_{R}^{\mathrm{sc}}(t^{\prime},k^{\prime}) & = & \theta\left(t^{\prime}\right)\frac{1}{2ic_{s}}\left(k_{x^{\prime}}^{2}+k_{\bot}^{2}\frac{1-c_{s}^{2}v^{2}}{1-v^{2}}\right)^{-1/2}\left(\text{e}^{i\omega_{+}t^{\prime}}-\text{e}^{i\omega_{-}t^{\prime}}\right),\label{Retardeed green from EOM app}\end{eqnarray}
 which coincides with the Green's function (\ref{Retarded Green in slow rocket app})
we calculated by applying the Lorentz transformation to the rest Green's
function in the case of slow motion. Note, however, that the results
differs for the case of fast moving spacecraft - compare (\ref{Retardeed green from EOM app})
and (\ref{Fast moving green correct app}). The function $G_{R}^{\mathrm{sc}}(t^{\prime},k^{\prime})$
from (\ref{Retardeed green from EOM app}) contains exponentially
growing modes for sufficiently high $k_{\bot}$, while correct way
of calculation gave us a sensible result (\ref{Fast moving green correct app})
- it contains only exponentially suppressed modes. This makes sense
because the late time solution approaches the free wave which do not
contain these high $k_{\bot}$.


\begin{thebibliography}{999}
\bibitem{Hawking}S.W.~Hawking, G.F.R.~Ellis, \emph{The Large scale structure of space-time},
Cambridge University Press, Cambridge, (1973).
\bibitem{Durrer}C.~Bonvin, C.~Caprini, R.~Durrer, Phys.Rev.Lett. \textbf{97}:081303,
(2006), [astro-ph/0606584].
\bibitem{Ellis}G.~F.~R.~Ellis, R.~Maartens, M.~MacCallum, gr-qc/0703121.
\bibitem{Gibbons}G. W. Gibbons, hep-th/0302199.
\bibitem{Zwanziger}G.~Velo, D.~Zwanziger, Phys.Rev.188:2218-2222, (1969);G.~Velo,
D.~Zwanziger, Phys.Rev. 186, 1337 - 1341 (1969)
\bibitem{Susskind}Y.~Aharonov, A.~Komar, L.~Susskind, Phys.~Rev.~\textbf{182}:1400-1403,
(1969).
\bibitem{Kleban}Antonio De Felice, Mark Hindmarsh, Mark Trodden, JCAP 0608:005, (2006),
[e-Print: astro-ph/0604154]; Gianluca Calcagni, Beatriz de Carlos,
Antonio De Felice, Nucl.Phys.B752:404-438,2006. [e-Print: hep-th/0604201];
A. Gruzinov, M. Kleban, hep-th/0612015;
\bibitem{Nima}A.~Adams, N.~Arkani-Hamed, S.~Dubovsky, A.~Nicolis, R.~Rattazzi,
JHEP \textbf{0610}:014, (2006), [hep-th/0602178].
\bibitem{Shore}G.M. Shore, e-Print: hep-th/0701185.
\bibitem{Recent}Timothy J. Hollowood, Graham M. Shore, arXiv:0707.2303; Timothy J.
Hollowood, Graham M. Shore, arXiv:0707.2302.
\bibitem{Blochinzev}D. Blochinzev, \emph{Space and time in microworld} (in Russian), Nauka,
1970.
\bibitem{Noncomutative}A. Hashimoto and N. Itzhaki, Phys.Rev. D63 (2001) 126004, [hep-th/0012093];
K. Landsteiner, E. Lopez and M.~H.~G. Tytgat, JHEP 0106 (2001) 055,
[hep-th/0104133]; Horatiu Nastase, hep-th/0601182.
\bibitem{Jacob}T.~Jacobson, D.~Mattingly, Phys.Rev.\textbf{D70}:024003 (2004),
e-Print: gr-qc/0402005.
\bibitem{Faster than gravity}I.T. Drummond and S.~J. Hathrell, Phys. Rev. D 22 (1980) 343; R.~D.~Daniels
and G.M. Shore, Nucl. Phys. B425 (1994) 634; R. D. Daniels and G.M.
Shore, Phys. Lett. B367 (1996) 75.
\bibitem{Ohkuma}Y. Ohkuwa, Progr. Theor. Phys. 65 1981 1058.
\bibitem{plates}K. Scharnhortst, Phys. Lett. B236 (1990) 354; G. Barton, Phys. Lett.
B237 (1990) 559; J.~I.~Lattorre, P.~Pascual, and R. Tarrach, Nucl.
Phys. B437 (1995) 60; S. Ben-Menahem, Phys. Lett. B250 (1990) 133.
\bibitem{UV}G.~Shore, Nucl. Phys. B460 (1996) 379, [gr-qc/9504041]; A.D.
Dolgov and I.B. Khriplovich, Zh. Eksp. Teor. Fiz. 58 (1983) 1153;
[English translation: Sov. Phys. JETP, 58 (1983) 671]; I.~B.
Khriplovich, Phys. Lett. B 346 (1995) 251.
\bibitem{Dolgov}A.~D.~Dolgov, I.~D.~Novikov, Phys.Lett. \textbf{B442}: 82 (1998),
[gr-qc/9807067].
\bibitem{Liberati}S.~Liberati, S.~Sonego, M.~Visser, Annals~Phys. \textbf{298}:167-185,
(2002), [gr-qc/0107091].
\bibitem{Chronology}S.W.~Hawking, Phys.Rev. \textbf{D46}, 603-611, (1992).
\bibitem{Godel}K.~Gödel, Rev.~Mod.~Phys. \textbf{21}, 447 (1949).
\bibitem{Gott}J.~R.~Gott, Phys.~Rev.~Lett. \textbf{66}, 1126 (1991).
\bibitem{Ori}A.~Ori, gr-qc/0701024.
\bibitem{otherCCC}A.~Ori, Phys.~Rev.~Lett.~\textbf{71}, 2517 (1993); Y.~Soen and
A.~Ori, Phys.~Rev.~\textbf{D54}, 4858 (1996); A.~Ori, Phys.~Rev.~Lett.~\textbf{95},
021101 (2005); R.L.~Mallett, Found.~Phys.~\textbf{33}, 1307 (2003);
W.~B.~Bonnor and B.R.~Steadman, Gen.~Rel.~Grav.~\textbf{37}:1833-1844
(2005); O.~Gron and S.~Johannesen, gr-qc/0703139.
\bibitem{wormhole}M.~S.~Morris, K.S.~Thorne and U.~Yurtsever, Phys.~Rev.~Lett.
\textbf{61}, 1446 (1988).
\bibitem{Kessence}C.~Armendariz-Picon, V.~F.~Mukhanov, P.~J.~Steinhardt, Phys.Rev.Lett.
\textbf{85}:4438-4441, (2000), {[}astro-ph/0004134{]}; C.~Armendariz-Picon,
V.F.~Mukhanov, P.J.~Steinhardt, Phys.~Rev.~\textbf{D63}:103510,
(2001), {[}astro-ph/0006373{]}.
\bibitem{Super infl}V.~F.~Mukhanov,~A.~Vikman,~JCAP~0602:004,~(2006), {[}astro-ph/0512066{]};
A.~Vikman, astro-ph/0606033.
\bibitem{GarMukh}J.~Garriga, V.~Mukhanov, Phys.~Lett. \textbf{B458}:219-225 (1999),
{[}hep-th/9904176{]}.
\bibitem{BH}E.~Babichev, V.~Mukhanov, A.~Vikman, JHEP \textbf{0609}:061,(2006),{[}hep-th/0604075{]};~E.~Babichev,
V.~Mukhanov, A.~Vikman, arXiv:0704.3301v1 {[}hep-th{]} .
\bibitem{Durrer2}C.~Bonvin, C.~Caprini, R.~Durrer, arXiv:0706.1538 {[}astro-ph{]};
\bibitem{Bruneton}J.-P.~Bruneton, Phys.~Rev.~\textbf{D75}:085013, (2007), {[}gr-qc/0607055{]};
J.-P.~Bruneton hep-th/0612113; J.-P.~Bruneton, G. Esposito-Farese,
arXiv:0705.4043v1 {[}gr-qc{]}.
\bibitem{KimIrSen}Jin U Kang, Vitaly Vanchurin, Sergei Winitzki, arXiv:0706.3994 {[}gr-qc{]};
\bibitem{Halo}C.~Armendariz-Picon, E.~A.~Lim, JCAP \textbf{0508}:007, (2005),
{[}astro-ph/0505207{]}.
\bibitem{Mukhanov BOOK}V.~Mukhanov, \emph{Physical foundations of cosmology}. Cambridge.
Univ. Pr. (2005)
\bibitem{Inst}James M. Cline, Sangyong Jeon, Guy D. Moore, Phys.Rev. D70 (2004)
043543, {[}hep-ph/0311312{]}; Sean M. Carroll, Mark Hoffman, Mark
Trodden, Phys.Rev. D68 (2003) 023509 {[}arXiv:astro-ph/0301273v2{]};
I. Ya. Aref'eva, I.V. Volovich, arXiv:hep-th/0612098v1; R. P. Woodard,
arXiv:astro-ph/0601672v2.
\bibitem{Rendall}A.D.~Rendall, Class.~Quant.~Grav.\textbf{~23}, 1557-1570, (2006),
{[}gr-qc/0511158{]}.
\bibitem{Novello}M.~Novello,~M.~Makler, L.~S.~Werneck, C.~A.~Romero, Phys.Rev.D71
(2005) 043515, {[}astro-ph/0501643{]}.
\bibitem{stringy causality}G.~W.~Gibbons, C.~A.~R. Herdeiro, Phys.~Rev.~D63:064006, (2001),
{[}hep-th/0008052{]}. G. Gibbons, Koji Hashimoto, Piljin Yi, JHEP
0209:061,2002, {[}hep-th/0209034{]}; G.~W.~Gibbons,~Class.~Quant.~Grav.20:S321-S346,~(2003),
{[}hep-th/0301117{]}; G.~W.~Gibbons, Rev.Mex.Fis.49S1:19-29, (2003),
{[}hep-th/0106059{]}.
\bibitem{Wald}R.~Wald, \emph{General relativity,} The University of Chicago Press,
(1984).
\bibitem{Leray}J.~Leray, \emph{Hyperbolic differential equations}, mimeographed
notes. Institute for Advanced Study, Princeton (1953).
\bibitem{Cauchy}H. Friedrich, A. Rendall, \emph{The Cauchy problem for the Einstein
equations}. In B. G. Schmidt (ed) \emph{Einstein's Field Equations
and Their Physical Implications}. Lecture Notes in Physics 540. Springer,
Berlin (2000), {[}gr-qc/0002074{]}; I. G. Petrovsky, \emph{Lectures
on partial differential equations}, Interscience Publishers, New York-London,
(1954).
\bibitem{BI}T. Damour, I. I. Kogan, Phys. Rev. D 66 (2002) 104024, {[}arXiv:hep-th/0206042{]};
D. Blas, C. Deffayet, J. Garriga, Class. Quant. Grav. 23 (2006) 1697,
{[}arXiv:hepth/ 0508163{]}; D.~Blas, C. Deffayet, J. Garriga, arXiv:0705.1982.
\bibitem{Bounce}Paolo Creminelli, Markus A. Luty, Alberto Nicolis, Leonardo Senatore,
JHEP 0612:080, (2006). e-Print: hep-th/0606090; Paolo Creminelli,
Leonardo Senatore, e-Print: hep-th/0702165; Evgeny I. Buchbinder,
Justin Khoury, Burt A. Ovrut, e-Print: hep-th/0702154; Evgeny I. Buchbinder,
Justin Khoury, Burt A. Ovrut, e-Print: arXiv:0706.3903 {[}hep-th{]}.
\bibitem{Ja}A.~Vikman, Phys.Rev.D71:023515, (2005), {[}astro-ph/0407107{]}.
\bibitem{Crossing}Robert R. Caldwell, Michael Doran, Phys.Rev.D72:043527, (2005), {[}astro-ph/0501104{]};
Anjan Ananda Sen, JCAP 0603:010, (2006), {[}astro-ph/0512406{]}; Luis
Raul Abramo, Nelson Pinto-Neto, Phys.Rev.D73:063522, (2006), {[}astro-ph/0511562{]};
Gong-Bo Zhao, Jun-Qing Xia, Mingzhe Li, Bo Feng, Xinmin Zhang, Phys.Rev.D72:123515,
(2005), {[}astro-ph/0507482{]}; Martin Kunz, Domenico Sapone, Phys.Rev.D74:123503,
(2006), {[}astro-ph/0609040{]}.
\bibitem{kinfl}C.~Armendariz-Picon, T.~Damour, V.~Mukhanov, Phys.Lett.~\textbf{B458}:209-218,
(1999), {[}hep-th/9904075{]}.
\bibitem{kself}Ferdinand Helmer, Sergei Winitzki, Phys.Rev.D74:063528,(2006); {[}gr-qc/0608019{]}.
\bibitem{DBI}M.~Alishahiha, E.~Silverstein, D.~Tong, Phys.~Rev.~D70:123505
(2004) {[}hep-th/0404084{]}; E.~Silverstein, D.~Tong, Phys.~Rev.~\textbf{D70}:
103505, (2004) {[}hep-th/0310221{]}.
\bibitem{Metric Redefinition}E. Fradkin and A. Tseytlin Phys. Lett. 15BB (1985), p. 316; David
J. Gross, Edward Witten, Nucl.Phys.B 277:1, (1986); B. Zwiebach, Phys.Lett.
156B (1985) 315; S. Deser and A.N. Redlich, Phys.Lett. 176B (1986)
350; A. A. Tseytlin, Phys.Lett. 176B (1986) 92; D.R.T. Jones and A.M.
Lowrence, Z.Phys. 42C (1989) 153.
\bibitem{Tolman}R. C. Tolman, \emph{The Theory of the Relativity of Motion}, Berkeley,
Univ. of California Press, (1917).
\bibitem{SuperGroup}D. Mugnai, A. Ranfagni, and R. Ruggeri, Phys. Rev. Lett. 84, 4830-4833
(2000); L. J. Wang, A. Kuzmich, and A. Dogarlu, Nature 406, 277-279
(2000).
\bibitem{Vladimirov}V.S.~Vladimirov, \emph{Equations of mathematical physics}, MIR (1984).
%
\bibitem{LL}L.~Landau, E.~Livshitz, \emph{Course of Theoretical Physics}, Vol.
2, \emph{The classical Theory of Fields,} Butterworth-Heinemann, (1987).
\bibitem{Visser}M.~Visser, Phys.Rev. \textbf{D46}:2445-2451 (1992) {[}hep-th/9203057{]}.
\bibitem{Moncrief}V.~Moncrief, Astrophysical Journal, Part 1, \textbf{235}, 1038-1046
(1980).
\bibitem{Born-Infel}M.~Born,~L.~Infeld,~Proc.~Roy.~Soc.~Lond.~\textbf{A144}:425-451,~(1934).
\bibitem{kdefect}E.~Babichev, Phys.Rev.D74:085004 (2006), {[}hep-th/0608071{]}; D.~Bazeia,
L.~Losano, R.~Menezes, J.C.R.E.~Oliveira, hep-th/0702052; X.~Jin,
X.~Li, D.~Liu, Class.Quant.Grav.\textbf{24}:2773-2780 (2007), arXiv:0704.1685
{[}gr-qc{]}.
\bibitem{Kofman}Gary N. Felder, Lev Kofman, Alexei Starobinsky, JHEP 0209:026, (2002),
{[}hep-th/0208019{]}.
\bibitem{Stockum}W.~J.~van~Stockum, Pros.R.Soc.Edinb. \textbf{57}, 135 (1937); F.
J. Tipler, Phys. Rev. D 9, 2203 (1974).
\bibitem{Analog}W.~G.~Unruh, Phys.Rev.Lett.46:1351-1353, (1981); Carlos Barcelo,
Stefano Liberati, Matt Visser, Living Rev.Rel.8:12, (2005), {[}gr-qc/0505065{]};
Matt Visser, Carlos Barcelo, Stefano Liberati, Gen.Rel.Grav.34:1719-1734,
(2002), {[}gr-qc/0111111{]}; \emph{Artificial black holes.} M. Novello,
(ed.) , M. Visser, (ed.), G. Volovik, (ed.), River Edge, USA: World
Scientific (2002).
\bibitem{Barbishov}B. Barbashov, N. Chernikov, \emph{Soviet Physics} JETP \textbf{5},
\textbf{50}, 1296-1308 (1966); B. Barbashov, N. Chernikov, \emph{Soviet
Physics} JETP \textbf{51}, 658 (1966).
\bibitem{Nonlinear}G. B. Whitham, \emph{Linear and nonlinear Waves} (Pure \& Applied
Mathematics), John Wiley \& Sons Inc; (1974)
\bibitem{Cuscuton}Niayesh Afshordi, Daniel J.H. Chung, Michael Doran, Ghazal Geshnizjani,
Phys.Rev.D75:123509, (2007), {[}astro-ph/0702002{]}; Niayesh Afshordi,
Daniel J.H. Chung, Ghazal Geshnizjani, Phys.Rev.D75:083513, (2007),
{[}hep-th/0609150{]}.
\end{thebibliography}
\end{document}